%% file: main.tex
\documentclass[journal]{IEEEtran}
\usepackage{xcolor}
\usepackage{cite}
\usepackage{amsmath,amssymb,amsfonts}
\usepackage{algorithmic}
\usepackage{mathtools}
\usepackage{textcomp}
\ifCLASSINFOpdf
  \usepackage[pdftex]{graphicx}

  \graphicspath{{figs/}}
\else
  \usepackage[dvips]{graphicx}
\fi

\usepackage{amsmath}
\usepackage{algorithmic}
\usepackage{array}

\ifCLASSOPTIONcompsoc
 \usepackage[caption=false,font=normalsize,labelfont=sf,textfont=sf]{subfig}
\else
 \usepackage[caption=false,font=footnotesize]{subfig}
\fi
\usepackage{dblfloatfix}
\usepackage[hyphens]{xurl}

\usepackage{cancel}
\usepackage{cite}

\usepackage{multirow}
\usepackage{threeparttable}

\begin{document}

\title{
 Receptive Field Regularization Techniques for Audio Classification and Tagging with Deep Convolutional Neural Networks}
%
%
%

\author{ \IEEEauthorblockN{Khaled Koutini\IEEEauthorrefmark{1},
        Hamid Eghbal-zadeh\IEEEauthorrefmark{1}\IEEEauthorrefmark{2},
       Gerhard Widmer\IEEEauthorrefmark{1}\IEEEauthorrefmark{2}}
       
       \IEEEauthorblockA{\IEEEauthorrefmark{1}Institute of Computational Perception (CP-JKU) \& \IEEEauthorrefmark{2}LIT Artificial Intelligence Lab,\\         
        Johannes Kepler University Linz, Austria\\
        khaled.koutini@jku.at\\ }

}

\markboth{IEEE/ACM TRANSACTIONS ON AUDIO, SPEECH, AND LANGUAGE PROCESSING}%
{Koutini \MakeLowercase{\textit{et al.}}}

\maketitle

\begin{abstract}
In this paper, we study the performance of variants of well-known Convolutional Neural Network (CNN) architectures on different audio tasks.
We show that tuning the Receptive Field (RF) of CNNs is crucial to their generalization. An insufficient RF limits the CNN's ability to fit the training data. 
In contrast, CNNs with an excessive RF tend to over-fit the training data and fail to generalize to unseen testing data. As state-of-the-art CNN architectures  -- in computer vision and other domains -- tend to go deeper in terms of number of layers, their RF size increases and therefore they degrade in performance in several audio classification and tagging tasks. We study well-known CNN architectures and how their building blocks affect their receptive field. We propose several systematic approaches to control the RF of CNNs and systematically test the resulting architectures on different audio classification and tagging tasks and datasets. 
The experiments show that regularizing the RF of CNNs using our proposed approaches can drastically improve the generalization of models, out-performing complex architectures and pre-trained models on larger datasets. 
The proposed CNNs achieve state-of-the-art results in multiple tasks, from \emph{acoustic scene classification} to \emph{emotion and theme detection in music} to \emph{instrument recognition}, as demonstrated by top ranks in several pertinent challenges (DCASE, MediaEval)\footnote{Code available: \url{https://github.com/kkoutini/cpjku_dcase20} }. 

\end{abstract}


%
\IEEEpeerreviewmaketitle

\section{Introduction}
\input{s1_intro}

\section{Related Work}
\input{s2_related_work}

\section{The Receptive Field of \\ Convolutional Neural Networks }
\input{s4_rf_cnns}

\section{Audio Classification and Tagging Tasks}
\input{s3_tasks_datasets}

\section{Controlling the Max RF}
\input{s5_control_maxrf_cnn}

\section{The Impact of the Max RF on Generalization}
\input{s6_result_maxrf_cnn}

\section{Controlling the Effective RF}

\input{s7_control_erf_cnn}
\section{The Impact of the ERF on Generalization}
\input{s8_result_erf_cnn}


\section{Conclusion}
\input{s9_conclusion}

\section*{Acknowledgment}
This work has been supported by the COMET-K2 Center of the Linz Center of Mechatronics (LCM) funded by the Austrian Federal Government and the Federal State of Upper Austria.
The LIT AI Lab is financed by the Federal State of
Upper Austria.

We thank the members of the Institute of Computational Perception for the useful discussions and feedback.

\bibliographystyle{IEEEtran}
\bibliography{refs}

\appendices

\section{Experimental Setup}

\input{a0_exp_setup}

\section{Common CNN Components and Their Effect on the RF}
\input{a1_arch}








\vfill
\vfill

\end{document}

%% file: s1_intro.tex
\IEEEPARstart{C}{\lowercase{o}}nvolutional Neural Networks (CNNs) have come to dominate
many domains~\cite{LIU201711survey}, including image classification~\cite{krizhevsky2012imagenetalex}, 
natural language processing~\cite{shen2014learning,kim-2014-convolutional,collobert2014deepcnnmultitask},
and speech recognition~\cite{hinton2012cnnspeach}.
In computer vision, CNNs have been shown to perform better with deeper architectures~\cite{simonyanVeryDeepConvolutional2014,heDeepResidualLearning2016,huangDenselyConnectedConvolutional2017}.
Following this trend, many researchers working on audio tagging and classification tasks have started
to employ such deep architectures from vision, by extracting image-like features such as
spectrograms~\cite{lee2009speccnn,eghbal-zadehCPJKUSubmissionsDCASE20162016,hersheyCNNArchitecturesLargescale2017}.

In this work, we show that using deeper architectures which have proven successful in vision tasks
do not necessarily guarantee good performance in audio tasks.
In particular, we show that deep CNNs designed for vision often have a large Receptive Field (RF)
that is not suitable for audio processing tasks, and that such models fail to generalize.
We provide solutions to this problem by regularizing the RF of different variants of deep architectures,
and we demonstrate that this massively improves results in various audio classification and tagging
tasks.

More specifically, we provide \textit{generic and systematic methods} for controlling
the RF of a CNN architecture. Using these, we show how well-known vision architectures
can be effectively adapted to the audio domain, by careful RF regularization.
We study the effect of RF regularization over the time and the frequency dimensions,
individually and jointly, and how the RF size over these dimensions affects the generalization of CNNs.
Generally, the experimental results show that the performance is more sensitive to the RF size
along the frequency dimension than over time.

As an additional important aspect, we differentiate between
(a) the \textit{Maximum Receptive Field} of a CNN, which can be calculated from its architecture,
and (b) the \textit{Effective Receptive Field}~\cite{luoUnderstandingEffectiveReceptive2016} (ERF)
of a trained CNN. We introduce a measure for the ERF and analyze the difference to the maximum RF. 
Finally, we introduce a novel method we call \emph{filter damping}, that enables us to control the ERF
of a CNN, without changing the architectural topology or the filters of the network. We show that
filter damping further helps improve generalization by adapting the ERF of models
to better fit the task at hand.

We demonstrate all this in large-scale experiments on three different audio classification
and tagging tasks--acoustic scene classification; emotion and theme detection in music; musical
instrument recognition--and using several datasets, testifying to
the generality of the problem and the proposed solution.\footnote{In addition to the tasks we  systematically study here, models built on top of our RF-Regularized CNNs have recently been shown
to excel also in several other audio classification problems, notably, device-invariant acoustic scene classification~\cite{Primus2019,Koutinitrrfcnns2019,Suh2020task1a,Koutini2020dcasesubmission},
audio tagging with noisy labels and minimal supervision~\cite{Fonseca2019,Koutinitrrfcnns2019},
open set acoustic scene classification~\cite{Lehner2019},
and low-complexity acoustic scene classification~\cite{Koutini2020dcasesubmission}.
}

As a general lesson, we propose to consider the RF of CNNs as an important hyper-parameter and design factor
that should be carefully tuned and investigated when adapting a CNN architecture (which has been designed
in other domains such as vision) to the audio world.

The remainder of the paper is structured as follows: after a brief discussion of the related work (Section~\ref{sec:related_work}), Section \ref{sec:rf_cnns} takes a closer look at the concepts of maximum and effective receptive field, and proposes ways to determine and measure them in a given model. 
We present the studied tasks and the data sets we have chosen for our investigations in Section~\ref{sec:tasks_datasets}.
The presentation of our proposed methods for systematically controlling RF in various deep CNN architectures, and the corresponding experimental investigations, are split into two strands: Sections \ref{sec:control_max_rf} and \ref{sec:results} focus on the maximum RF, while Sections \ref{sec:control_erf} and \ref{sec:result_erf}  discuss the ERF. We explain the setup we use to perform the experiments in Appendix~\ref{app:exp_setup}.  
Finally, Appendix~\ref{app:arch:design:components:maxrf}  presents an analysis of the influence of particular components of CNNs on the RF.

%% file: s2_related_work.tex
\label{sec:related_work}

With the power of feature learning, CNNs have revolutionized the field of machine vision and image processing.
CNNs can learn hierarchies of features, from low-level concepts such as edges to high-level representations of shapes and objects, which contribute to their superior performance.

Several deep CNN architectures have been proposed in the vision domain that have set new milestones.
The VGG architectures~\cite{simonyanVeryDeepConvolutional2014} introduced an effective structure of convolutional and pooling layers that outperformed other models at the time.
However, due to the vanishing gradient issue, they were limited in terms of depth (number of processing layers).
To address this issue, \textit{ResNet}~\cite{heDeepResidualLearning2016} and \textit{DenseNet}~\cite{huangDenselyConnectedConvolutional2017} architectures were proposed that tackled the vanishing gradient problem by incorporating skip connections to allow better gradient flow through layers and permit the use of very deep architectures.

Despite the success of very deep architectures in the vision world, the audio domain is still dominated by VGG-like architectures~\cite{eghbal-zadehCPJKUSubmissionsDCASE20162016,lehnerClassifyingShortAcoustic2017,DorferDCASE2018task1,SakashitaDCASE2018task1}.
For instance, for the task of \textit{acoustic scene classification}, Eghbal-zadeh et al.~\cite{eghbal-zadehCPJKUSubmissionsDCASE20162016} adapted the VGG architecture from computer vision to use spectrograms as input. Their proposed architecture in combination with an i-vector-based system achieved the top rank in DCASE 2016.
Hershey et al.~\cite{hersheyCNNArchitecturesLargescale2017} compared various well-known image recognition CNN architectures on a large-scale dataset of 70M audio clips from YouTube.
They showed that on such a large dataset, very deep CNNs such as ResNet-50 can perform very well. 
However, deep architectures have shown less success on smaller datasets~\cite{Zhao2017} compared to more shallow architectures such as VGG~\cite{mesaros2017dcase}.
As a result, many state-of-the-art acoustic scene classification systems still incorporate shallow CNN architectures~\cite{lehnerClassifyingShortAcoustic2017,DorferDCASE2018task1,SakashitaDCASE2018task1}, since these architectures have been shown to generalize better given the size of the available datasets.


Several neurophysiological studies have focused on understanding the ability of humans and animals to tune their cortical Spectro-Temporal Receptive Fields
(STRFs) in order to selectively focus on target sounds, while minimizing the irrelevant acoustics and noise background~\cite{carlin2015frameworkspeeachdetectionrf,fritz2007auditory,bajo2010focusing,shamma2014adaptive}. Building on such studies, Carlin and Elhilali~\cite{carlin2015frameworkspeeachdetectionrf} trained a  Gaussian Mixture Model on features obtained from both the initial and adapted STRFs. They showed that an ensemble of adapted STRFs achieves better performance in detecting speech, in the presence of noise. 
These results suggest that adapting the RF is important, also in biological auditory systems.


Several efforts in the literature have focused on adapting CNNs to audio and music tasks.
Pons et al.~\cite{ponsExperimentingMusicallyMotivated2016} examined the effect of filter shapes in shallow CNN architectures in music classification tasks and proposed to change the shape of convolutional filters in order to restrict CNNs to learn either temporal or frequency dependencies in the data, which is more relevant to the characteristics of musical signals.

Regarding \textit{theme and emotion detection in music}, another one of our 
test domains (see Section~\ref{sec:mtg:jemando}), it is again CNNs that are the most popular models, as evidenced by the submissions to the MediaEval 2019 benchmark~\cite{bogdanov2019mtg,koutini2019emotion,sukhavasi2019music_cnnselfattention,amiriparian2019emotion}.
The benchmark baseline~\cite{bogdanov2019mtg} uses a VGG-based architecture. 
Sukhavasi and Adapa~\cite{sukhavasi2019music_cnnselfattention} use MobileNetV2~\cite{sandler2018mobilenetv2} with self-attention~\cite{vaswani2017attention} to capture temporal relations in musical signals.
Amiriparian et al~\cite{amiriparian2019emotion} use pre-trained CNNs as feature extractors.
These CNNs are pre-trained on Audioset~\cite{audioset2017Gemmeke} as well ImageNet~\cite{dengImagenetLargescaleHierarchical2009} and their extracted features were further used with recurrent NNs as classifiers. 
Again, we show in this work that RF-regularized CNNs can outperform the aforementioned complex approaches.

Finally, in \textit{musical instrument recognition} (which we discuss in Section~\ref{sec:openmic}), a common approach is to use CNNs that are pre-trained on large datasets, and use them as feature extractors to produce embeddings~\cite{audioset2017Gemmeke,hersheyCNNArchitecturesLargescale2017}, which are then fed into classifiers that map them to the instrument classes~\cite{humphrey2018openmic,GururaniSL19openmoc_attn,amir2020openmic}.
Humphrey et al.~\cite{humphrey2018openmic} use random forests to classify the embeddings, while Gururani et al.~\cite{GururaniSL19openmoc_attn} study fully connected networks, recurrent
neural networks, and an attention-based model. 
Anhari~\cite{amir2020openmic} uses an LSTM~\cite{hochreiter1997lstm} with an attention layer to model the instruments. 
As in the previous cases, we show in the present paper that a \textit{simple RF-regularized CNN} trained only on OpenMIC can outperform such complex approaches and architectures, including those using models pre-trained on large datasets.

%% file: s4_rf_cnns.tex
\label{sec:rf_cnns}
In this section, we explain two important concepts in CNNs, namely the maximum RF and the Effective RF, which--as shown in later sections--are important for the generalization capabilities of CNNs in audio tasks.
\subsection{The Maximum Receptive Field of CNNs}
\label{sec:rf_cnns:mrf}

A neuron in a convolutional layer is affected only by a part of the layer's input, as opposed to fully-connected layers, where each neuron takes the whole output of the previous layer as input. 
In other words, each neuron in a convolutional layer has a specific `field of view'. The input of the layer outside of this \textit{field of view} cannot alter the neuron's activation.
This \textit{field of view} is known as the Receptive Field (\textit{RF}).

In this study, we keep our focus on the receptive field over the spatial dimensions of the input (e.g, frequency and time, in the case of spectrograms). In the general case, the output activation of a neuron in a convolutional layer depends on the spatial dimensions of the input and on the channel dimension that represents the number of input feature maps.

We focus on fully-convolutional networks since they have demonstrated superior performance compared to CNNs that incorporate fully-connected output layers~\cite{DorferDCASE2018task1,chen2019dcaseasc}.
Moreover, a fully-connected layer is equivalent to a convolutional layer with no padding and a filter equal in size to the spatial size of the last feature map.    
In this section, we start with basic CNN architectures. We detail the more advanced architecture modules in  Appendix~\ref{app:arch:design:components:maxrf}.

Within a convolutional layer, the spatial RF of a neuron is determined by its filter size: the larger the filter, the bigger the region of the input it can ``see''. 
The filter size determines the RF of a convolutional layer over the layer input. However, the RF of a layer with respect to the \textit{network input} (the spectrogram)--that is, what the neuron ``sees'' of the input--is dependent on the number and configuration of the previous layers as well. 
The filter size, the stride, and the dilation of all the previous layers influence the RF of a neuron with respect to~the input.

In a CNN, the size $RF_n$ of the receptive field of a unit from layer $n$ to the network input, along a particular input dimension, can be calculated as:
\begin{align}
\label{eq:calc_maxrf}
RF_n &= RF_{n-1}+(k_n-1) \cdot S_n , \\
\nonumber S_n &=S_{n-1} \cdot s_n 
\end{align}
where $s_n$, $k_n$ are stride and kernel size of layer $n$, respectively, and $S_n$ is the cumulative stride from layer $n$ to the input layer.

\subsection{The Effective Receptive Field of CNNs}
\label{sec:rf_cnns:erf}

The previous section described the \textit{maximum} RF of a neuron, that is, the region of the input that is connected to a neuron. 
However, parts of the input may have minimal or no influence on the neuron's activation, depending on the learned parameters of the model.
The set of input pixels or units that effectively influence a neuron is called its \textit{Effective Receptive Field (ERF).}
Luo et al.~\cite{luoUnderstandingEffectiveReceptive2016} show that neurons are more influenced by input pixels around the center of their max RF, since this region of the input has more paths to the neuron on both the forward and the backward pass. 
They show that the effective receptive field is different from the theoretical maximum receptive field. 
Further, they 
propose a method for computing the Effective Receptive Field (ERF) of a trained CNN, by back-propagating a gradient signal from an output neuron through the network to the input. This method reveals the regions of the input that affect the output the most.

In order to compute the ERF in models trained on audio spectrograms,
we follow the approach proposed by Luo et al.~\cite{luoUnderstandingEffectiveReceptive2016}.
We back-propagate a gradient signal from the output of the penultimate layer\footnote{The layer before the final layer, which is usually the one before removing the spatial information with global pooling.} to the inputs.
Luo et al.~\cite{luoUnderstandingEffectiveReceptive2016} visualize the ERF using random data as input, while we use an unseen test set to calculate the gradients with respect to the input.
Therefore, we can analyze the relationship between the ERF and the generalization on this unseen test set.

\begin{figure}[t]
\centering
\includegraphics[width=3.5in]{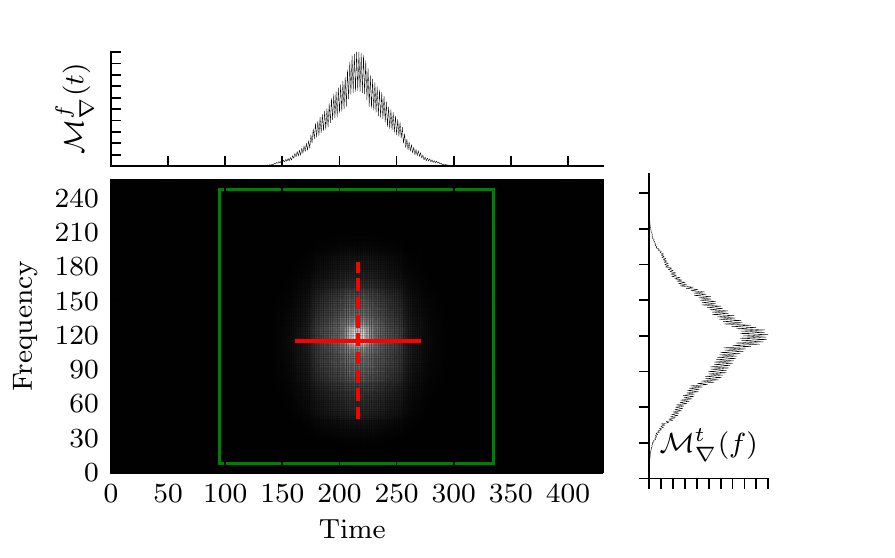}
\caption{The ERF in white, the maximum RF in green, and our estimates $E_t$ and $E_f$ in red, for a ResNet with a Maximum RF of $241 \times 241$ trained on the DCASE`18  dataset.}
\label{fig:erf_visual}
\end{figure}

We now propose an objective measure of the ERF size in terms of two scalars $E_t$ and $E_f$ indicating the ERF size over the time and frequency dimensions, respectively.
We define $\mathcal{M}_\nabla{(t,f)}$ as the mean of the gradients (over the test set) on an input pixel specified by the pixel coordinates $(t,f)$, where $t$ refers to the coordinate of the time axis, and $f$ represents the pixel coordinate of the frequency axis in the input spectrogram.
Let $T$ and $F$ be the dimension, in pixels, of the input along time and frequency, and $\mathcal{M}^{t}_\nabla({f})=\sum_{t=1}^{T}\mathcal{M}_\nabla{(t,f)}$, $\mathcal{M}^{f}_\nabla({t})=\sum_{f=1}^{F}\mathcal{M}_\nabla{(t,f)}$ be the marginal sum of the mean gradients over frequency and time dimensions, respectively, as visualized in Figure~\ref{fig:erf_visual}.
The ERF size over the \textit{time} dimension, $E_t$, is defined as $E_t= 4 \cdot \sigma_t$, where $\sigma_t$ is the standard deviation of the marginal sum $\mathcal{M}^{f}_\nabla({t})$.
We estimate the center of the ERF $\mu_t$ and standard deviation $\sigma_t$ on the time dimension as shown in ~\eqref{eq:calc_erf_center}:


\begin{align}    
\nonumber \mu_t &= \frac{ \sum_{t=1}^{T} t \cdot \mathcal{M}^{f}_\nabla({t}) }{\sum_{t=1}^{T} \mathcal{M}^{f}_\nabla({t})}, \\
\sigma_t &= \sqrt{ \frac{ \sum_{t=1}^{T}(t - \mu_t)^2 \cdot \mathcal{M}^{f}_\nabla({t}) }{\sum_{t=1}^{T} \mathcal{M}^{f}_\nabla({t})}}
\label{eq:calc_erf_center}
\end{align}



Luo et al.~\cite{luoUnderstandingEffectiveReceptive2016} showed that the gradient used in the ERF calculation reaches the maximum value in the center of the Max RF, and the gradient magnitude decays away from the center in a squared exponential way.
This can also be observed in our visualisation of the ERF in Figure~\ref{fig:erf_visual}.
Given this shape, we can conclude that the range $[\mu_t -2\sigma_t,\mu_t +2\sigma_t]$ with the length of $E_t$, contains most of the impactful pixels, as $ E[\frac{\sum_{t=\mu_t -2\sigma_t}^{t=\mu_t +2\sigma_t} {\mathcal{M}^{f}_\nabla({t})}}{\sum_t {\mathcal{M}^{f}_\nabla({t})}}] \approx  0.9545$.
Therefore, we propose $E_t$ as measure of the ERF, as it represent the size of the range that contains the most impactful pixels in the time dimension.

Analogously, we estimate $E_f$, the effective ERF size along the frequency axis, by calculating $\mu_f$ and $\sigma_f$ in a similar fashion to Eq.~\ref{eq:calc_erf_center}, and defining $E_f= 4 \cdot \sigma_f$.

%% file: s3_tasks_datasets.tex
\label{sec:tasks_datasets}
In this section, we introduce the tasks and the respective datasets to be used in our empirical analysis.
The tasks, intended to cover a variety of challenges, 
are acoustic scene classification, emotion and theme prediction in music, and musical instrument recognition.

\subsection{Acoustic Scene Classification}

The task in Acoustic Scene Classification (ASC) is to classify an acoustic scene (such as city center and park) using a short (10-30 seconds) audio recording. 
This task has various applications such as content-based multimedia information retrieval, context-aware smart devices and monitoring systems. 
The evaluation measure commonly used for this task is classification accuracy, as the common benchmark datasets are generally balanced in terms of samples from each class.
Accuracy is  the official evaluation measure for ASC in the \textit{IEEE AASP Challenge on Detection and Classification of Acoustic Scenes and Events} (DCASE\footnote{\url{http://dcase.community}}).
The following two recent ASC benchmark datasets are used.

\subsubsection{TAU Urban Acoustic Scenes 2018}
This dataset was released for the DCASE 2018 Challenge, Task 1~\cite{MesarosDCASE2018T1}, and contains 10-second audio clips recorded from 10 different acoustic scenes. 
The training data consists of around 1000 minutes of audio. The evaluation (test) data consists of 420 minutes and includes audio clips recorded in  cities and countries not included in the training set. Each audio segment must be classified into one out of 10 possible acoustic scenes. We refer to this dataset as \textit{DCASE`18}.

\subsubsection{TAU Urban Acoustic Scenes 2019}
In DCASE 2019 Challenge Task 1~\cite{Mesaros2019}, the DCASE'18 acoustic scenes classification dataset was extended by adding audio recorded in 6 new cities. We refer to this dataset as \textit{DCASE`19}.

\subsection{Emotion and Theme Detection in Music}
\label{sec:mtg:jemando}
The goal of this tagging task is to automatically recognize the emotions and `themes' (general concepts or tags such as `love', `film', `space') conveyed in music recordings, which has many applications in Music Information Retrieval (MIR). 
We use the dataset released in the \textit{Emotion and Theme Recognition in Music} Task at the MediaEval-2019 Challenge\footnote{\url{https://multimediaeval.github.io/2019-Emotion-and-Theme-Recognition-in-Music-Task/}}.
This is a subset of the MTG-Jamendo dataset~\cite{bogdanov2019mtg}, which we refer to as \textit{MediaEval-MTG-Jamendo dataset} throughout this paper. The subset consists of 9,949 and 4,231 recordings for training and  testing, respectively. Each recording is tagged with one or more out of 57 possible labels.

We use Precision-Recall Area Under Curve (PR-AUC) as the performance measure for this task.
PR-AUC accounts for different decision threshold values, and was the official evaluation measure in the MediaEval challenge.

\subsection{Musical Instrument Recognition}
\label{sec:openmic}
Instrument recognition is the task of recognizing the presence of musical instruments in an audio clip. 
We focus our study on  polyphonic instrument recognition, where multiple instruments are present in the signal. 
This is considered to be significantly more difficult~\cite{Lostanlen2018instrument,GururaniSL19openmoc_attn,GururaniSL18instrument,Han2016instrumentidentification} than monophonic  instrument recognition, where the task is to recognize the instrument from a note or an audio signal containing a single instrument.

\textit{OpenMIC-2018}~\cite{humphrey2018openmic} is a freely available dataset of 20,000 audio clips from the Free Music Archive~\cite{Defferrard2017fda}. 
Each clip is 10 seconds long and is annotated with tags indicating the presence or absence of 20 possible instruments.
Because multiple instruments can appear in a single clip, this is a multi-label instrument recognition problem.  The dataset is known to be noisy because of the particular way the ground truth tags were collected via crowd-sourcing, leading to a potentially large fraction of missing labels and annotator disagreements.

%% file: s5_control_maxrf_cnn.tex
\label{sec:control_max_rf}

Equation~\eqref{eq:calc_maxrf} in Section~\ref{sec:rf_cnns} shows that we can control the RF of a basic CNN by changing the number of layers $N$ and the filter size $k_n$ and stride $s_n$ of a layer $n$. In Appendix~\ref{app:arch:design:components:maxrf}, we investigate some of the widely-used components that are commonly used in famous CNN architectures, and discuss how they relate to the change in the Max RF. 
In this section, we propose a method to systematically control the RF of two CNN architectures, and explain our proposed RF-regularized CNNs in Section~\ref{sec:rfreg:cns}. These two architectures are the basis of our experiments. 
In Section~\ref{sec:control:max:2d}, we extend our method to control the Max RF on each dimension independently, which allows us to investigate the impact of the Max RF in time or frequency separately.

\subsection{RF-Regularized CNNs}
\label{sec:rfreg:cns}

\subsubsection{ResNets}
\label{sec:control_maxrf:resnet}

ResNet~\cite{heDeepResidualLearning2016} is a very well-known CNN architecture that enables the architecture to be very deep while maintaining the gradient flow, by incorporating residual connections between layers.
ResNet variants exhibited state-of-the-art performance in many computer vision tasks~\cite{He2016preact}. 
However, as explained in Section~\ref{sec:rf_cnns}, the Max RF of CNNs tends to grow as the networks go deeper, by stacking more layers. A larger RF--as we will show--hinders the CNN's ability to generalize in audio tasks. 

We control the Max RF of a ResNet by changing the filters of a number of its convolutional layers from $3 \times 3$ to $1 \times 1$, since $1 \times 1$ filters do not increase the Max RF of the network (see Equation~\eqref{eq:calc_maxrf}). 
We call the resulting architecture family \emph{CP\_ResNet}. 
The Max RF of a \emph{CP\_ResNet} is controlled by a hyper-parameter $\rho$, which determines the number of $3 \times 3$ and  $1 \times 1$ filters. 
The resulting \emph{CP\_ResNet} architecture consist of a $5 \times 5$ convolutional layer followed by a $3 \times 3$ layer, a $1 \times 1$ layer and $2 \times 2$ pooling. 
This is then followed by $\rho$ layers with $3 \times 3$ filters and $(21-\rho)$  layers with $1 \times 1$, as specified in Table~\ref{tab:resnet_configs} and Equation~\ref{eqn:rfset}:
\begin{equation}
  \label{eqn:rfset}
 x_k = 
     \begin{cases}
        3 &\quad\text{if }k\le \rho \\
       1 &\quad\text{if } k > \rho \\
     \end{cases}
\end{equation}
 
Values of $\rho$ ranging from 0 to 21 result in networks with a Max RF ranging from $23 \times 23$ to $583 \times 583$, as shown in Table~\ref{tab:rho_max_rf}.

\begin{table}[t]
\caption{CP\_ResNet architectures family}
\begin{center}
\begin{tabular}{|c|c|}
\hline
\textbf{RB Number}&\textbf{RB Config} \\
\hline
&Input $ 5 \times 5$ stride=$2$ 
\\
\hline

1&$3 \times 3$, $ 1 \times 1$, P\\
2  & $ x_1 \times x_1$,  $ x_2 \times x_2$, P  \\
3  & $ x_3 \times x_3$,  $ x_4 \times x_4$  \\
4    & $ x_5 \times x_5$,  $ x_6 \times x_6$, P   \\
5&$x_7 \times x_7$, $ x_8 \times x_8$  \\
6 &$ x_9 \times x_9$, $ x_{10} \times x_{10}$  \\
7 &$ x_{11} \times x_{11}$, $ x_{12} \times x_{12}$ \\
8  &$ x_{13} \times x_{13}$, $ x_{14} \times x_{14}$  \\
9 &$ x_{15} \times x_{15}$, $ x_{16} \times x_{16}$  \\
10&$ x_{17} \times x_{17}$, $ x_{18} \times x_{18}$  \\
11 &$ x_{19} \times x_{19}$, $ x_{20} \times x_{20}$  \\
12  &$ x_{21} \times x_{21}$, $ x_{22} \times x_{22}$  \\
\hline
\multicolumn{2}{l}{RB: Residual Block, P: $ 2 \times 2$ max pooling.}\\
\multicolumn{2}{l}{$x_k \in \{ 1 , 3 \}$: is determined by $\rho$ (Equation~\ref{eqn:rfset})}\\
\multicolumn{2}{l}{ of the network. Number of channels per RB:}\\
\multicolumn{2}{l}{128 for RBs 1-4; 256 for RBs 5-8; 512 for RBs 9-12.} 
\end{tabular}
\label{tab:resnet_configs}
\end{center}
\end{table}

\begin{table}[t]
\caption{
Mapping $\rho$ values to the maximum RF (applies to both CP\_ResNet and CP\_DenseNet).
}
\begin{center}
\begin{tabular}{|c|c||c|c|}
\hline
\textbf{ $\rho$ value}&\textbf{Max RF}&\textbf{ $\rho$ value}&\textbf{Max RF} \\ \hline
0 & $ 23  \times  23  $
&
1 & $ 31  \times  31  $
\\ \hline
2 & $ 39  \times  39  $
&
3 & $ 55  \times  55  $
\\ \hline
4 & $ 71  \times  71  $
&
5 & $ 87  \times  87  $
\\ \hline
6 & $ 103  \times  103  $
&
7 & $ 135  \times  135  $
\\ \hline
8 & $ 167  \times  167  $
&
9 & $ 199  \times  199  $
\\ \hline
10 & $ 231  \times  231  $
&
11 & $ 263  \times  263  $
\\ \hline
12 & $ 295  \times  295  $
&
13 & $ 327  \times  327  $
\\ \hline
14 & $ 359  \times  359  $
&
15 & $ 391  \times  391  $
\\ \hline
16 & $ 423  \times  423  $
&
17 & $ 455  \times  455  $
\\ \hline
18 & $ 487  \times  487  $
&
19 & $ 519  \times  519  $
\\ \hline
20 & $ 551  \times  551  $
&
21 & $ 583  \times  583  $
\\ \hline
\hline
\end{tabular}
\label{tab:rho_max_rf}
\end{center}
\end{table}

\begin{table}[t]
\caption{CP\_DenseNet architectures family}
\begin{center}
\begin{tabular}{|c|}
\hline
\textbf{Convolutional Layers Config} \\
\hline
Input $ 5 \times 5$ stride=$2$ 
\\
\hline

B $3 \times 3$ C , B $ 1 \times 1$ C, B P\\
 B $ x_1 \times x_1$ C, B  $ x_2 \times x_2$ C,  P  \\
 B $ x_3 \times x_3$ C, B  $ x_4 \times x_4$  \\
 B $ x_5 \times x_5$ C, B  $ x_6 \times x_6$ C,  P   \\
B $x_7 \times x_7$ C, B $ x_8 \times x_8$ C 
B $ x_9 \times x_9$ C, B $ x_{10} \times x_{10}$ C \\
B $ x_{11} \times x_{11}$ C, B $ x_{12} \times x_{12}$ C
B $ x_{13} \times x_{13}$ C, B $ x_{14} \times x_{14}$ C \\
B $ x_{15} \times x_{15}$ C, B $ x_{16} \times x_{16}$C  
B $ x_{17} \times x_{17}$ C, B $ x_{18} \times x_{18}$ C \\
B $ x_{19} \times x_{19}$ C, B $ x_{20} \times x_{20}$ C 
B $ x_{21} \times x_{21}$ C, B $ x_{22} \times x_{22}$ C \\
\hline
\multicolumn{1}{l}{B: $ 1 \times 1$ convolution, P: $ 2 \times 2$ max pooling.}\\
\multicolumn{1}{l}{C: Concatenating the input of the DenseLayer.}\\
\multicolumn{1}{l}{A Dense layer consists of B $x \times x$ C.}\\
\multicolumn{1}{l}{$x_k \in \{ 1 , 3 \}$: is determined by $\rho$ (Equation ~\ref{eqn:rfset})}\\
\end{tabular}
\label{tab:densenet_configs}
\end{center}
\end{table}

\subsubsection{DenseNets~\cite{huangDenselyConnectedConvolutional2017}}
\label{sec:control_maxrf:densenet}
 are another successful architecture with state-of-the-art performance in many computer vision tasks. 
Similar to ResNets, DenseNets use skip connections to 
allow the back-propagation of  a stronger learning signal to the first layers. 

\emph{Dense layers} are the building block of DenseNets~\cite{huangDenselyConnectedConvolutional2017}.  
Each Dense layer consists of two convolutional layers: the first projects the input using $1 \times 1$ filters and therefore does not affect the Max RF. 
The second convolutional layer uses $3 \times 3$ filters that increase the Max RF.
 The \emph{growth rate} is a hyper-parameter that determines the number of channels in the output of each Dense layer. This output will be concatenated to the layer's input and passed  as input to the next layers.

We construct \emph{CP\_Dense} by choosing the filter sizes in the \emph{Dense layers} according to a parameter $\rho$, which
controls the number of $3 \times 3$ convolutions in the Dense layers that are changed to $1 \times 1$, in a similar fashion in CP\_ResNet. 
Table~\ref{tab:densenet_configs} and Equation~\eqref{eqn:rfset} define the final architecture for a $\rho$ value. 
Table~\ref{tab:rho_max_rf} shows the Max RF for both CP\_DenseNet and CP\_ResNet.

The output of each Dense layer is connected to all the following Dense layers, and therefore contributes to increasing the Max RF of the network. However, the growth rate does not change the Max RF, since it only affects the number of filters in each Dense layer.

\subsubsection{VGG baseline}
We use the shallow VGG-like architecture proposed by~\cite{DorferDCASE2018task1} as a baseline. 
The network has a max RF of $135 \times 135$ pixels.

\subsection{Controlling RF Dimensions Separately}
\label{sec:control:max:2d}

In order to analyze the impact of the Max RF over a single dimension (time or frequency), we need to control this dimension independently.
We accomplish this by extending the proposed method, using two independent hyper-parameters $\rho_f$ and  $\rho_t$ to control the Max RF over the frequency and the time dimension, respectively. 
We replace \emph{CP\_ResNet} filters (Table~\ref{tab:resnet_configs}) in the $k^{th}$ layer from $x_k \times x_k$ with  $x_k^f \times x_k^t$, where $x_k^f$ and $x_k^t$ are controlled by $\rho_f$ and  $\rho_t$, respectively, as shown in Equation~\ref{eqn:rfset:2d}:

\begin{equation}
  \label{eqn:rfset:2d}
 x_k^d = 
     \begin{cases}
        3 &\quad\text{if }k\le \rho_d \\
       1 &\quad\text{if } k > \rho_d \\
     \end{cases}
\end{equation}
where $d$ is either the time or the frequency  dimension.

%% file: s6_result_maxrf_cnn.tex
\label{sec:results}

In this section, we evaluate and analyze the prediction and generalization capabilities of CNNs under various Max RF settings. 
In Section~\ref{sub:sec:maxrf:overfitting}, we show that CNNs with large RFs show signs of overfitting. We evaluate the CNNs on the studied tasks in Section~\ref{subsec:results:tasks}. Further, in Section~\ref{sec:results:2d}, we analyze the impact of the RF size on each dimension independently.

\subsection{Overfitting the Training Data }
\label{sub:sec:maxrf:overfitting}

\begin{figure}[h]
\centering
\includegraphics[width=3.5in]{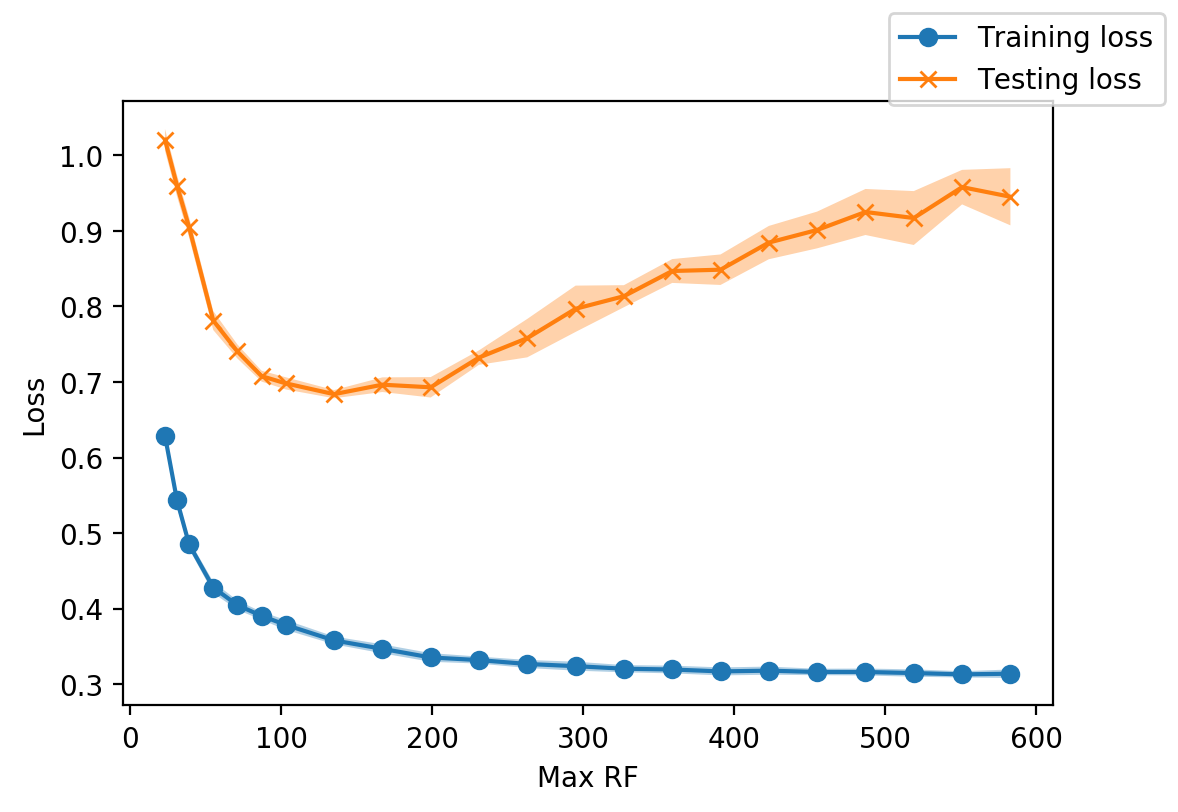}
\caption{Training loss vs.~testing loss of a CNN with different Max RFs. The increase of the testing loss with larger Max RF indicates overfitting. 
}
\label{fig:train_vs_test_by_maxrf}
\end{figure}

We first look at an experiment that indicates that although CNNs with larger RF fit the training data better, they fail to generalize to the test data after a certain limit. 
We analyze the impact of changing the Max RF on the training and testing loss of a CP\_ResNet on an acoustic scene classification task, using the DCASE'18 dataset (Section~\ref{sec:tasks_datasets}).
We train variants of CP\_ResNet (Section~\ref{sec:control_maxrf:resnet}) with different Max RF settings.
Figure~\ref{fig:train_vs_test_by_maxrf} shows the training and testing loss of the model, as the Max RF increases. 
While we can see that increasing the max RF always results in a decrease in the training loss, this is not the case for the testing loss. As the max RF continues to grow larger than 200, the testing loss starts to rise. This indicates that while the CNN's capability to fit the training data has been increased by growing the Max RF, the model starts to overfit if the Max RF becomes very large.
Figure~\ref{fig:train_vs_test_by_maxrf_pooling} in Appendix B
 shows the same effect when increasing the Max RF without changing the number of parameters of the CNN.

\subsection{Performance on Audio Classification and Tagging Tasks}
\label{subsec:results:tasks}

\begin{figure}[t]
\centering
\includegraphics[width=3.5in]{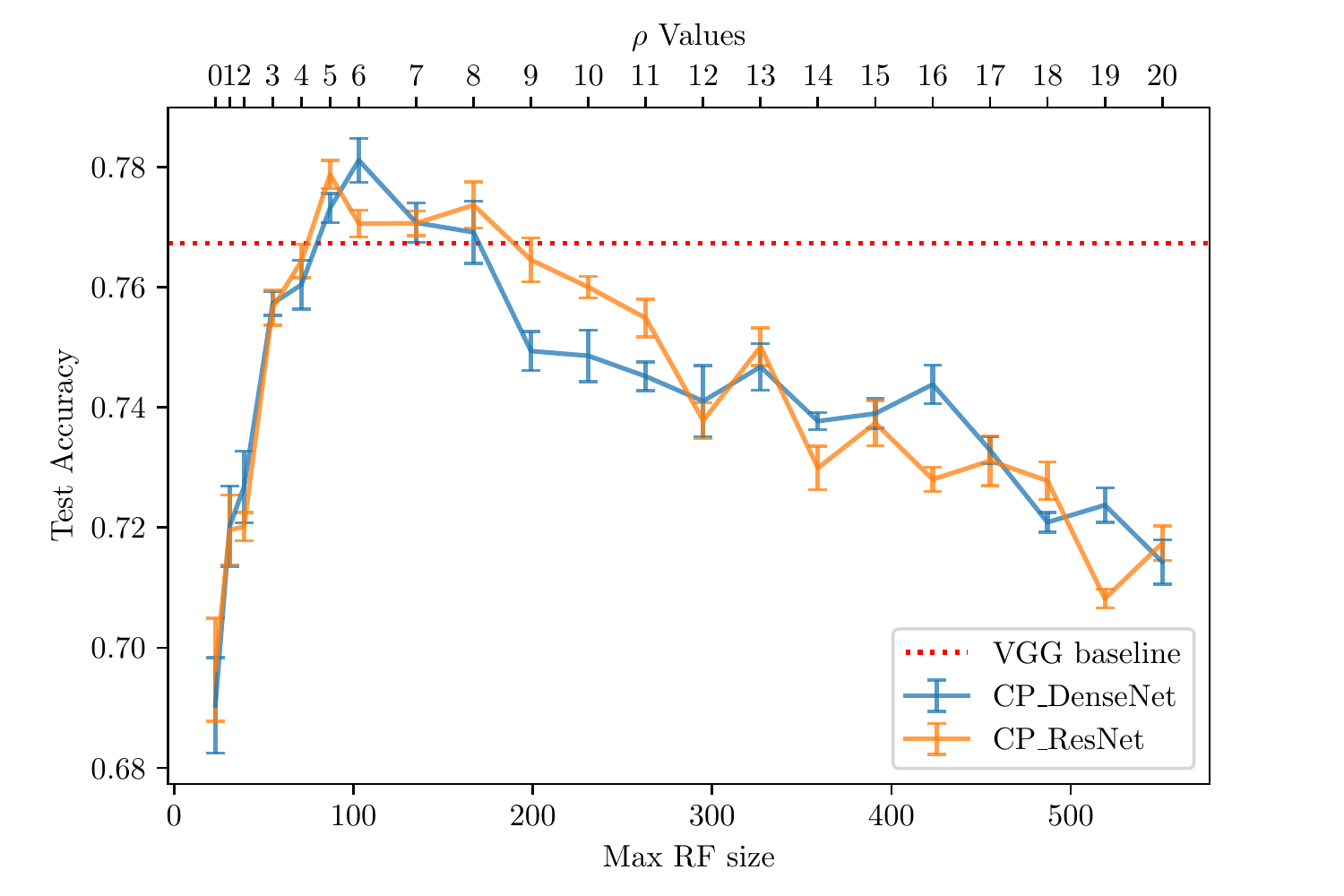}
\caption{Testing accuracy of CP\_Dense and CP\_ResNet on DCASE`18. The Max RF size of the network over both time and frequency dimensions (x-axis) ranges from $23 \times 23$ to $551 \times 551$}
\label{fig:dcase18:main:acc}
\end{figure}

In this section, we present the results of training CP\_ResNet and CP\_DenseNet on the three studied tasks introduced in Section~\ref{sec:tasks_datasets}. \\


\subsubsection{Acoustic Scene Classification}
Figure~\ref{fig:dcase18:main:acc} shows the prediction accuracy of CP\_ResNet and CP\_DenseNet models on the test part of the DCASE`18 dataset, with a \emph{growth rate} of $64$.
The plots show that both architectures achieve the best performance in the range of approximately $100 \times 100$ to $200 \times 200$ pixels. 
When the Max RF goes below this range, performance degrades.
This can be explained by the fact that the network underfits the training data as shown in Figures~\ref{fig:train_vs_test_by_maxrf} and~\ref{fig:train_vs_test_by_maxrf_pooling}. 
In contrast, as the Max RF goes beyond the optimal range, the networks show signs of overfitting 
and therefore degrade in performance.
It is worth noting that the average accuracy of CP\_ResNet$_{\rho=5}$ $0.779$ lies within the 95\%-confidence interval of the VGG baseline $[0.747,0.780]$, while the average performance of CP\_DenseNet$_{\rho=6}$ $0.781$ lies outside this interval.

\begin{table}[t]
\caption{Accuracy Comparison of CP\_ResNet with the state of the art on the DCASE`19 Dataset}
\begin{center}
\begin{tabular}{llll}
Method             & Seen Cities &  Unseen Cities & Overall \\
\hline
Chen et al.~\cite{chen2019dcaseasc}          & \textbf{86.7} \%               & 77.9 \%                  & \textbf{85.2} \%  \\
CP\_ResNet CV     & 84.8 \%               & \textbf{78.5} \%                  & 83.7 \%  \\
CP\_ResNet Single & 84.2 \%               & 75.8 \%                  & 82.8 \% \\
Seo et al.~\cite{Hyeji2019dcaseasc} & 83.8   \%               & 76.5 \%                  &  82.5 \% \\
\end{tabular}
\label{tab:res:asc:sota}
\end{center}
\end{table}

Table~\ref{tab:res:asc:sota} offers a performance comparison between the state-of-the-art models in acoustic scene classification, based on the submissions to the DCASE 2019 challenge\footnote{The full table of results can be found on the challenge's website~\url{http://dcase.community/challenge2019/task-acoustic-scene-classification-results-a}}\cite{Mesaros2019}. 
We show the results on both the unseen cities (no recordings from these cities were present in the training set), 
seen cities, and the total accuracy. 
Chen et al.~\cite{chen2019dcaseasc} achieved the first place using an ensemble of 9 different models (including different CNNs and RNNs), trained on different types of input features. 
They utilize a complex data augmentation scheme using Auxiliary Classifier Generative Adversarial Network (ACGAN)~\cite{OdenaOS17cgan} and Conditional Variational Autoencoder (CVAE)~\cite{SohnLY15cvae} and adversarial training as domain adaptation techniques, for a better generalization to unseen cities.
Seo et al.~\cite{Hyeji2019dcaseasc} achieved third place in the challenge using an ensemble of 8 models consists of 3 types of CNNs trained on various types of pre-processed input features. 
CP\_ResNet-based submissions achieved second place in the challenge. 
In addition, a \emph{single} model CP\_ResNet 
trained on the whole training set (``CP\_ResNet Single'' in the table) reaches comparable performance.

Further results for the CP\_DenseNet 
with different \emph{growth rates} can be found in
Figure~\ref{fig:dcase18:dilated:grouped:growth:acc} in Appendix B.
Additionally, the class-wise performance of CP\_ResNet on DCASE`19 and a comparison with the SOTA are provided in Table~\ref{tab:res:asc:sota:perclass}.
We refer the reader to~\cite{Koutini2019Receptive,Koutinitrrfcnns2019} for detailed results on DCASE 2016, DCASE 2017 and DCASE 2019 datasets. \\




\subsubsection{Emotion and Theme Recognition in Music}

\begin{figure}[t]
\centering
\includegraphics[width=3.5in]{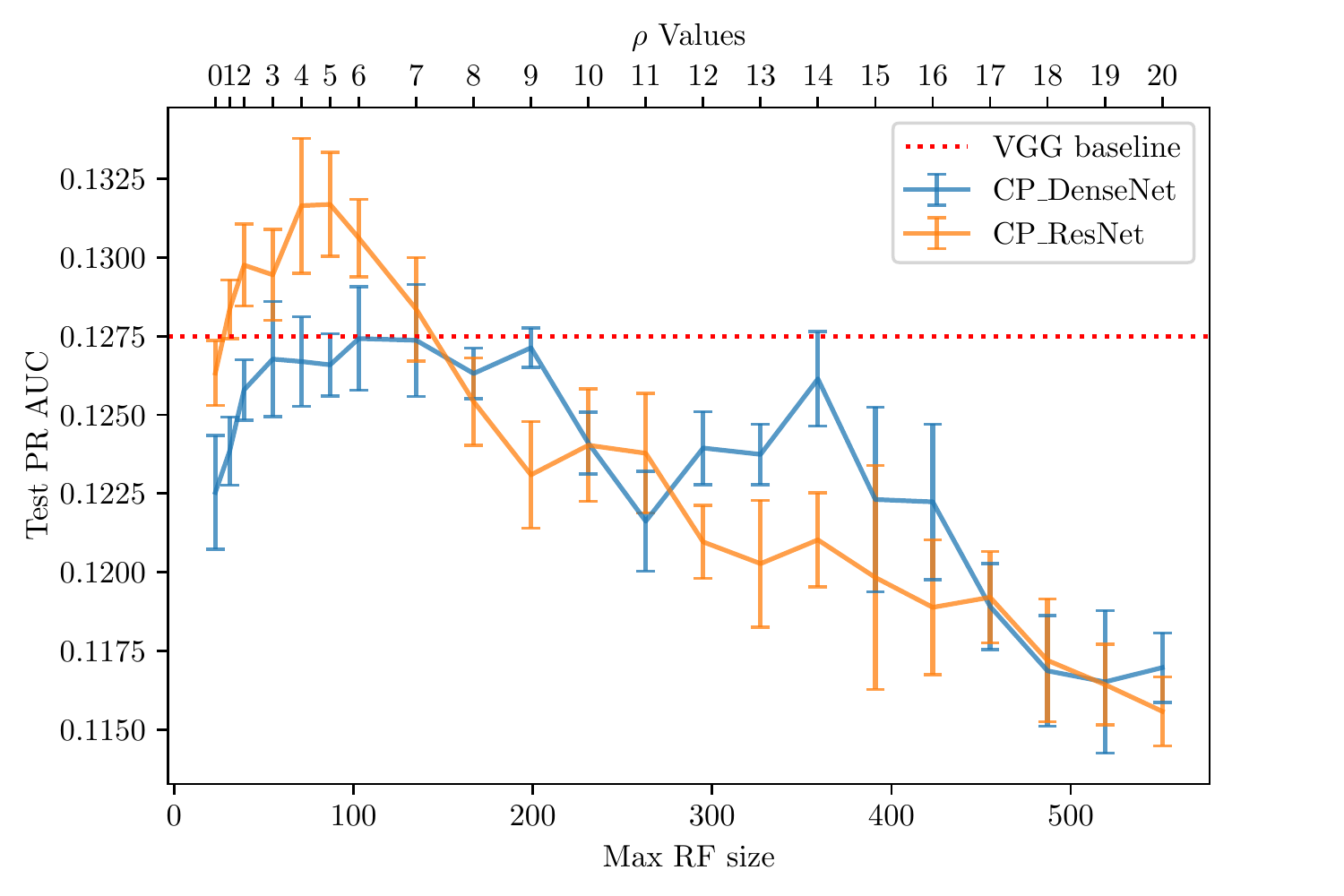}
\caption{PR\_AUC of CP\_Dense and CP\_ResNet on  MediaEval-MTG-Jamendo test set. The Max RF size of the network over both time and frequency dimensions (x-axis) ranges from $23 \times 23$ to $551 \times 551$}
\label{fig:mediaeval:main:acc}
\end{figure}

Figure~\ref{fig:mediaeval:main:acc} shows the Precision-Recall Area Under Curve (PR\_AUC)  for CP\_ResNet and CP\_DenseNet 
trained on the MediaEval-MTG-Jamendo Dataset (see Section~\ref{sec:mtg:jemando}).
The plots show that the optimal Max RF range for this tagging task in CP\_ResNet is approximately between $50 \time 50 $ pixels and  $100 \time 100$.
CP\_DenseNet has a wider Max RF range where it maintains its best performance before degrading in larger RF experiments. 


\begin{table}[t]
\caption{Performance Comparison of the state of the art on the test set of MediaEval-MTG-Jamendo Dataset}
\begin{center}
\begin{tabular}{ll}
Method            &  PR\_AUC  
\\
\hline
RF-Reg. CNNs Ensemble*         &  \textbf{  .1546	}        
\\
CP\_ResNet$_{\rho=5}\dagger$ &      .1312$ \pm  .0017 $   
\\
CP\_DenseNet$_{\rho=9}\dagger$ &     .1271$ \pm.0006 $   
\\
SS CP\_ResNet$_{\rho=9}\dagger$ &     .1395$ \pm.0013 $  
\\
Sukhavasi and Adapa~\cite{sukhavasi2019music_cnnselfattention}*  &    .1259    
\\
Amiriparian et al~\cite{amiriparian2019emotion}*   &    .1175      
\\
 VGG Baseline (ours)~\cite{DorferDCASE2018task1}  &      .1275$ \pm  .0015 $   
\\
 Baseline~\cite{bogdanov2019mtg}*  &      .1077     
\\
\hline
\multicolumn{2}{l}{ $\dagger$: single model, not submitted to the challenge.}\\
\multicolumn{2}{l}{*: Results taken from the challenge.}\\
\multicolumn{2}{l}{SS: Shake-Shake}\\
\end{tabular}
\label{tab:res:mediaEval:sota}
\end{center}
\end{table}

Table~\ref{tab:res:mediaEval:sota} offers a comparison of the state-of-the-art methods on emotion and theme recognition in music, based on the models submitted to the MediaEval 2019 Challenge.\footnote{We refer to the challenge website for the full list of the submitted methods~\url{https://multimediaeval.github.io/2019-Emotion-and-Theme-Recognition-in-Music-Task/results}}  
Sukhavasi and Adapa~\cite{sukhavasi2019music_cnnselfattention} used MobileNetV2~\cite{sandler2018mobilenetv2} with self-attention~\cite{vaswani2017attention}.
Amiriparian et al~\cite{amiriparian2019emotion} used pre-trained CNNs and RNNs.
Our submission, under ``RF-regularized CNNs Ensemble''~\cite{koutini2019emotion}, consisted in of an ensemble of CP\_ResNet, CP\_FAResNet, and Shake-Shake regularized CNNs as explained in Appendix~\ref{app:arch:design:components:maxrf}.
As can be seen in Table~\ref{tab:res:mediaEval:sota}, our model achieved first place in the challenge by a considerable margin, in terms of PR\_AUC.
Rows 2-4 in the table refer to the three individual component models of our ensemble, which were not submitted to the challenge but still seem to surpass the other submissions.
These results strongly suggest that RF-regularized CNNs are capable of achieving top performance,
without pre-training, attention, or recurrent modules. \\

\subsubsection{Musical Instrument Tagging}
\begin{figure}[t]
\centering
\includegraphics[width=3.5in]{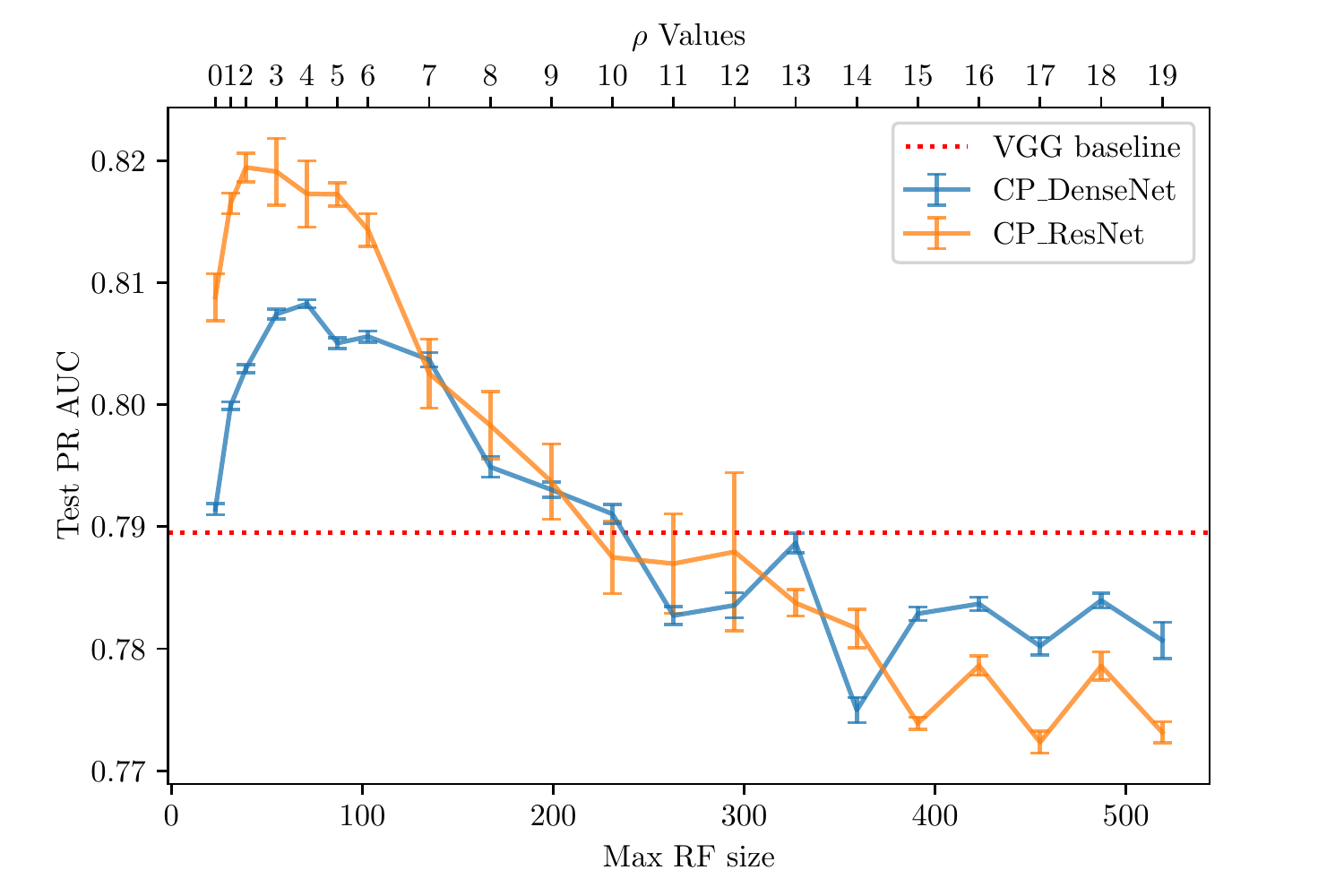}
\caption{PR-AUC of CP\_Dense and CP\_ResNet on OpenMIC dataset. The Max RF size of the network over both time and frequency dimensions (x-axis) ranges from $23 \times 23$ to $551 \times 551$}
\label{fig:openmic:main:prauc}
\end{figure}

\begin{figure}[t]
\centering
\includegraphics[width=3.5in]{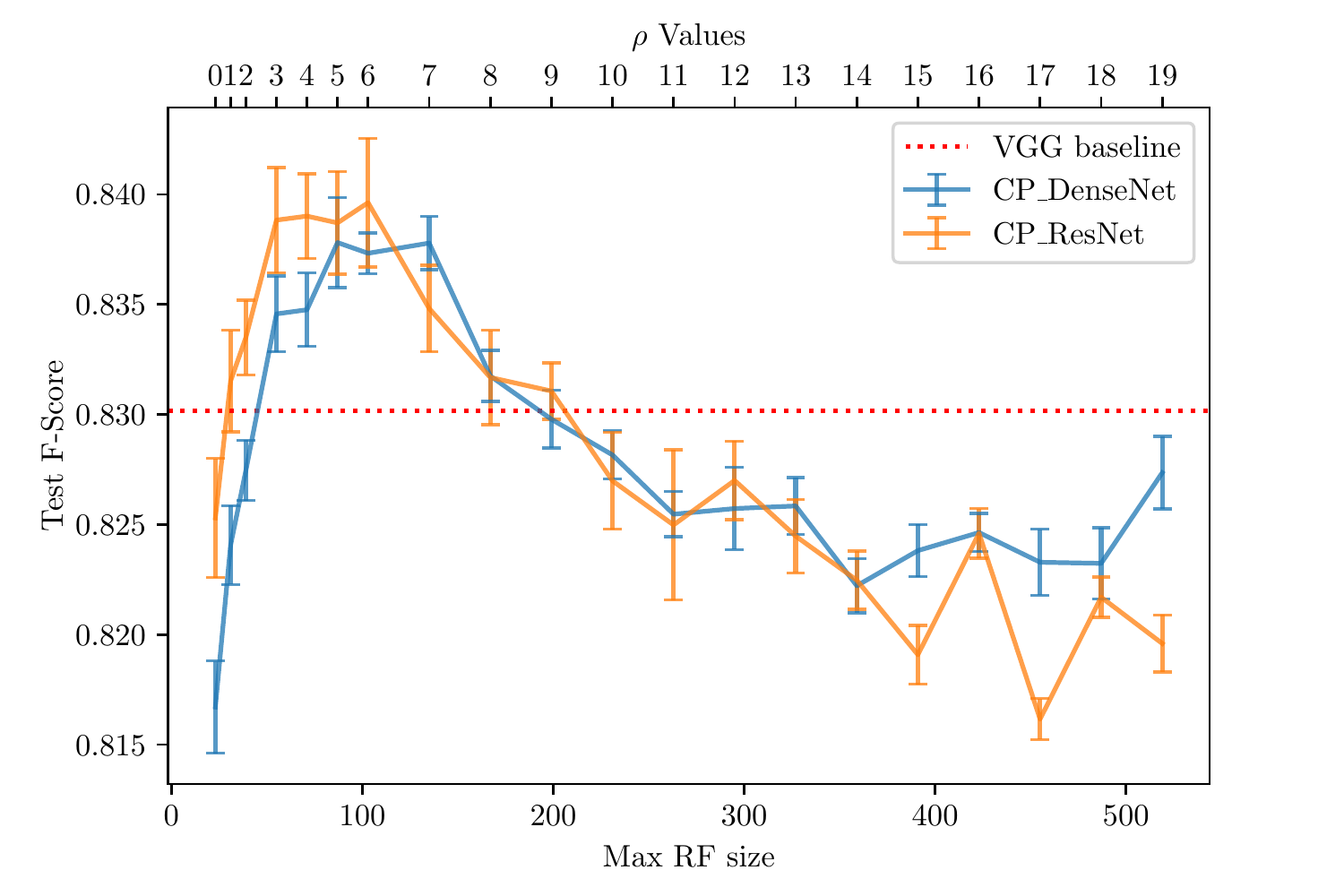}
\caption{F-Score of CP\_Dense and CP\_ResNet on OpenMIC dataset. The Max RF size of the network over both time and frequency dimensions (x-axis) ranges from $23 \times 23$ to $551 \times 551$}
\label{fig:openmic:main:fscore}
\end{figure}

Figures~\ref{fig:openmic:main:prauc} and \ref{fig:openmic:main:fscore} show PR-AUC and F-score on the test part of the OpenMIC~\cite{humphrey2018openmic} dataset. 
The figures show that the optimal Max RF for both architectures is approximately between $50 \time 50$ and  $100 \time 100$ pixels.
Similar to the results in the previous tasks, we can see that the performance of both architecture families degrades rapidly as the Max RF of the networks grows.

\begin{table}[t]
\caption{Performance  comparison of  the state-of-the-art on the test set of OpenMIC (cited methods are sorted by publication date). }
\begin{threeparttable}
\centering
\tabcolsep=0.11cm
\begin{tabular}{llll}
Method            &  PR\_AUC   & F-score\tnote{b} & F-score\tnote{a}  \\
\hline
CP\_ResNet$_{\rho=3}$ &      .819$ \pm.001 $    &  .809$ \pm.003 $  &   .847$ \pm.002 $  \\
SS CP\_ResNet$_{\rho=7}$ &      .831$ \pm.000 $ &    .822$ \pm.001$   &  .855$ \pm  .002 $ \\
Castel-Branco ~\cite{Branco2020openmic} &   .701         	            &    -    & -  \\
Anhari~\cite{amir2020openmic} &   -        	            &    -    &  .83\tnote{c}  \\

Gururani~\cite{GururaniSL19openmoc_attn}      &    -    &  .81\tnote{c} & - \\
 Baseline~\cite{humphrey2018openmic} &      .795  &   .785  &  .826 \\
VGG Baseline (ours)~\cite{DorferDCASE2018task1}  &      .789$ \pm.003 $   &   .801$ \pm.001$ &  .83$ \pm.002$ \\
 
\\
\hline
\end{tabular}
\begin{tablenotes}
\item[a] Classical F-score as used in~\cite{amir2020openmic,humphrey2018openmic}
\item[b]Average of the F-score of the positive and negative class  per instrument as proposed in  ~\cite{GururaniSL19openmoc_attn} 
\item[c] Indicates that the number is approximately read from the figures in the referenced papers
\item[] SS: applying Shake-Shake regularization.
\end{tablenotes}
\end{threeparttable}
\label{tab:res:openmic:sota}
\end{table}

Table~\ref{tab:res:openmic:sota} shows the state-of-the-art in instrument tagging in polyphonic music on the OpenMIC dataset.
All of the previous state-of-the-art approaches~\cite{amir2020openmic,GururaniSL19openmoc_attn,humphrey2018openmic} use CNNs pre-trained on the Audioset dataset~\cite{audioset2017Gemmeke}. 
They then train an audio tagging model using the CNN's embedding as input, 
which in~\cite{humphrey2018openmic} is a random-forest, in~\cite{GururaniSL19openmoc_attn} is an attention module, and in ~\cite{amir2020openmic} is an RNN with attention. 
Comparing these to our CP\_ResNets, we again see that simple RF-regularized CNNs can achieve at least comparable performances without using larger external datasets, pre-training, or stacking additional classification models. 

\subsection{Studying the Influence of the Time and Frequency Dimensions Independently }
\label{sec:results:2d}

\begin{figure}[t]
\centering
\includegraphics[width=3.5in]{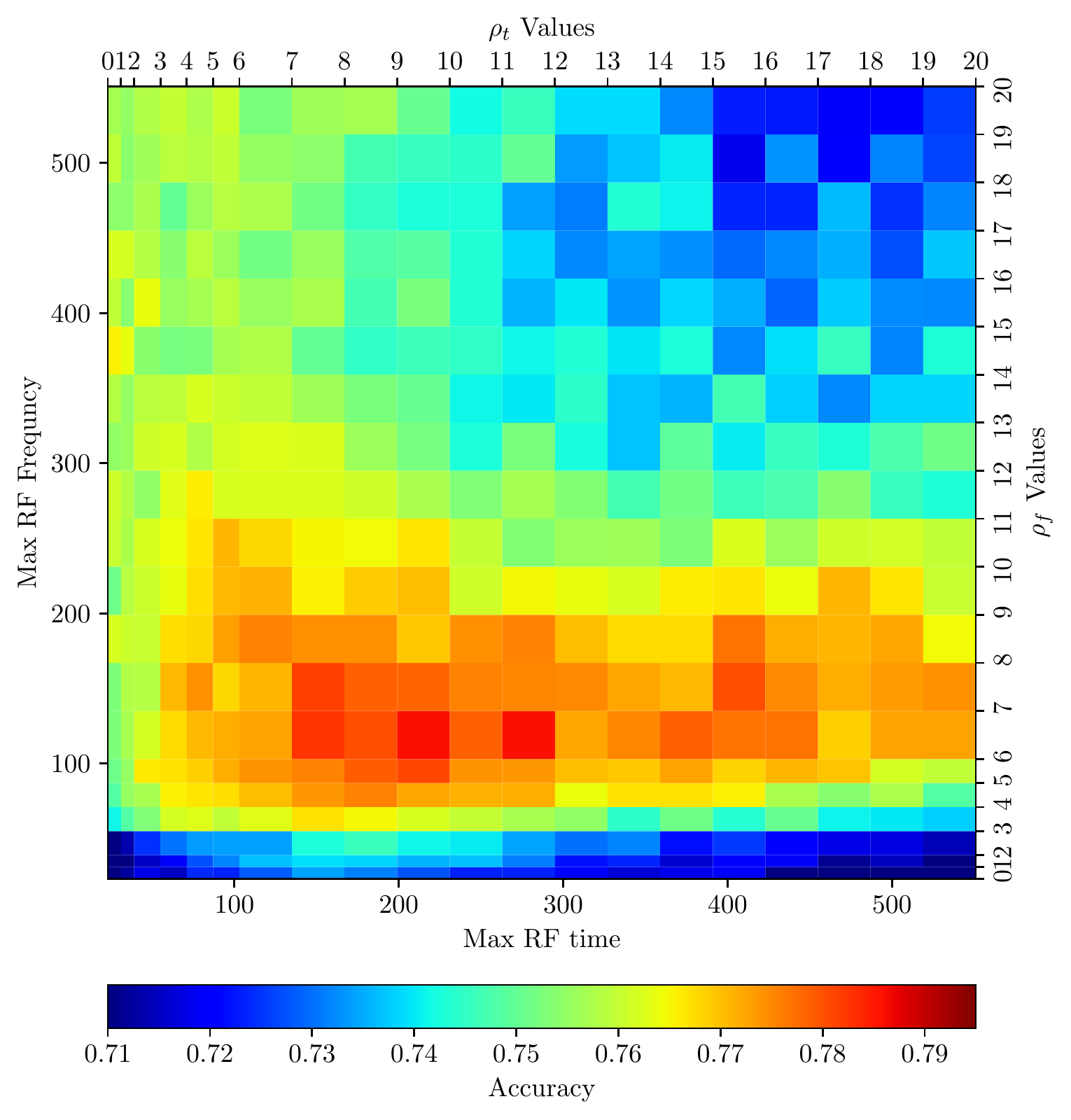}
\caption{Test set accuracy of different variants of CP\_ResNet with changing Max-RF over time (x-axis) and frequency (y-axis). Max-RF ranges from $23 \times 23$ to $551 \times 551$}
\label{fig:dcase18:2d_acc:time_freq}
\end{figure}

\begin{figure}[t]
\centering
\includegraphics[width=3.5in]{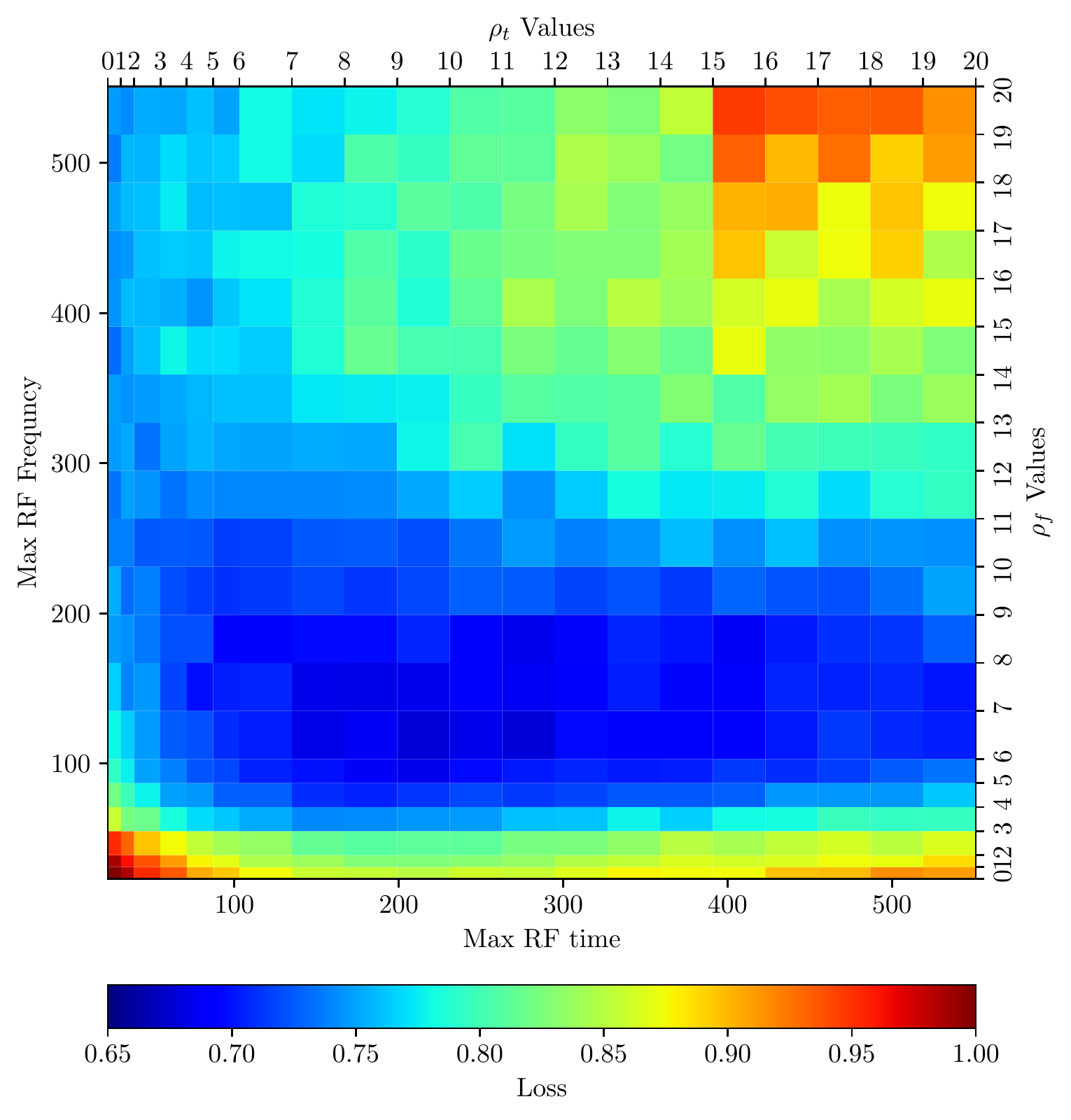}
\caption{Loss on test set of different variants of CP\_ResNet with changing Max-RF over time (x-axis) and frequency (y-axis). Max-RF ranges from $23 \times 23$ to $551 \times 551$}
\label{fig:dcase18:2d_loss:time_freq}
\end{figure}

The previous experiments showed the impact of changing the Max RF over both the frequency and the time dimension of the spectrograms.  
In this section, we analyze the impact of the Max RF in the two dimensions independently. 
We study the modified version of CP\_ResNet (as explained in Section~\ref{sec:control:max:2d}) by testing instances of CP\_ResNet with various $\rho_f$ and $\rho_t$ values on the DCASE'18  dataset.

Figures~\ref{fig:dcase18:2d_acc:time_freq} and~\ref{fig:dcase18:2d_loss:time_freq} show the test set accuracy and loss for different Max RF values over each of the dimensions. 
We can see that a Max RF approximately between $100$ and $200$ over the frequency, correlates with the highest performance (similar to our findings in Figure~\ref{fig:dcase18:main:acc}). Furthermore, as long as Max RF over the frequency dimension is within this range, increasing the Max RF over the time dimension has only minor effects on the performance.

The two figures also show that too small a Max RF over frequency (below $55$ pixels) generally results in poor performance. However, this can be partly compensated for
by choosing a large RF over time (see the bottom strip in the figures) -- at least in this particular task.
 

%% file: s7_control_erf_cnn.tex
\label{sec:control_erf}



In this section, we investigate restricting the ERF while keeping other factors of a CNN  -- such as architecture, topology, number of parameters --  constant. 
We achieve this by modifying a CNN with a specific Max RF. We decay the influence of the input, the further the input is from the center of the RF. 
This process does not affect the Max RF of the CNN but decreases the size of its ERF (Section~\ref{sec:rf_cnns:erf}).
We call the resulting networks \emph{damped} CNNs. 
We damp all the filters of a convolutional layer by performing an element-wise multiplication of the kernel with a constant matrix $C \in \mathbb{R}^{T\times F} $ before apply the convolution operator. We call $C$ the \emph{damping matrix}. 
The damping matrix has the same spatial shape as the filter and works by weakening the effect of spatially-outermost weights of the filter.  In other words,
we replace every convolution operation $O_n  = W_n \star Z_{n-1} + B_n$ with $O_n = (W_n \odot C_n) \star Z_{n-1} + B_n$
, where $\star$ is the convolution operator, $\odot$ is the Hadamard product, $Z_{n-1}$ is the output of the previous layer, $W_n$ and  $B_n$ are the weight and bias of the layer $n$.

     

For our experiments, we choose a damping matrix $ C \in \mathbb{R}^{T\times F} $ whose elements are given by the following equation:
\begin{align}    
\label{eq:damping}
c_{t,f} = \left( 1-m_t \frac{\mathopen| t-T/2 \mathclose|}{T/2}\right) \left(1-m_f\frac{\mathopen|f-F/2\mathclose|}{F/2}\right)
\end{align}
where $t \in \{0,\dots,T-1\} $, $f \in \{0,\dots,F-1\}$ are the matrix element indices, $T$, $F$ are the filter sizes over time and frequency respectively.  $m_f$, $m_t$ are hyper-parameters that control the damping in each dimension.
Multiplying the weights with the damping matrix $W_n \odot C_n$  is equivalent to scaling the weights (corresponding to the outer regions of the Max RF) to smaller values, and scaling their gradients down without changing any other factor in the training setup. Therefore, damping can be seen as an inductive bias hindering the network from fitting the outer regions of its max RF during training.
Moreover, damping uses element-wise multiplication with the weights during training which has linear complexity with respect to the number of parameters and negligible computationally compared to the matrix multiplications and convolution operations. After training, the weights can be updated by multiplying with the damping matrix, and therefore no additional computational complexity is added at inference time.

Similar to the experiments on controlling the Max RF, we investigate the effect of damping in both dimensions simultaneously, or separately.
\subsubsection{Damping on Both Dimensions}
We set $m_t=m_f=0.9$, which means that $C$ decays linearly away from the center along each dimension, scaling down the effects of input pixels further from the center in each dimension.

\subsubsection{Damping on the Frequency Dimension}
In these experiments, we set $m_t=0, m_f=0.9$, which results in a linear decay of $C$ away from the center only over the frequency dimension. This scales down the effects of the input pixels with distance from the center in the frequency dimension while keeping the influence of the input regardless of the distance from the center over the time dimension. 

\subsubsection{Damping on the Time Dimension}
This similar to the previous section but done on the time dimension.

%% file: s8_result_erf_cnn.tex
\label{sec:result_erf}
In this section, we present the results of restricting the ERF of the CP\_ResNet.
\subsection{Performance on Audio Classification and Tagging Tasks}


\begin{figure}[t]
\centering
\includegraphics[width=3.7in]{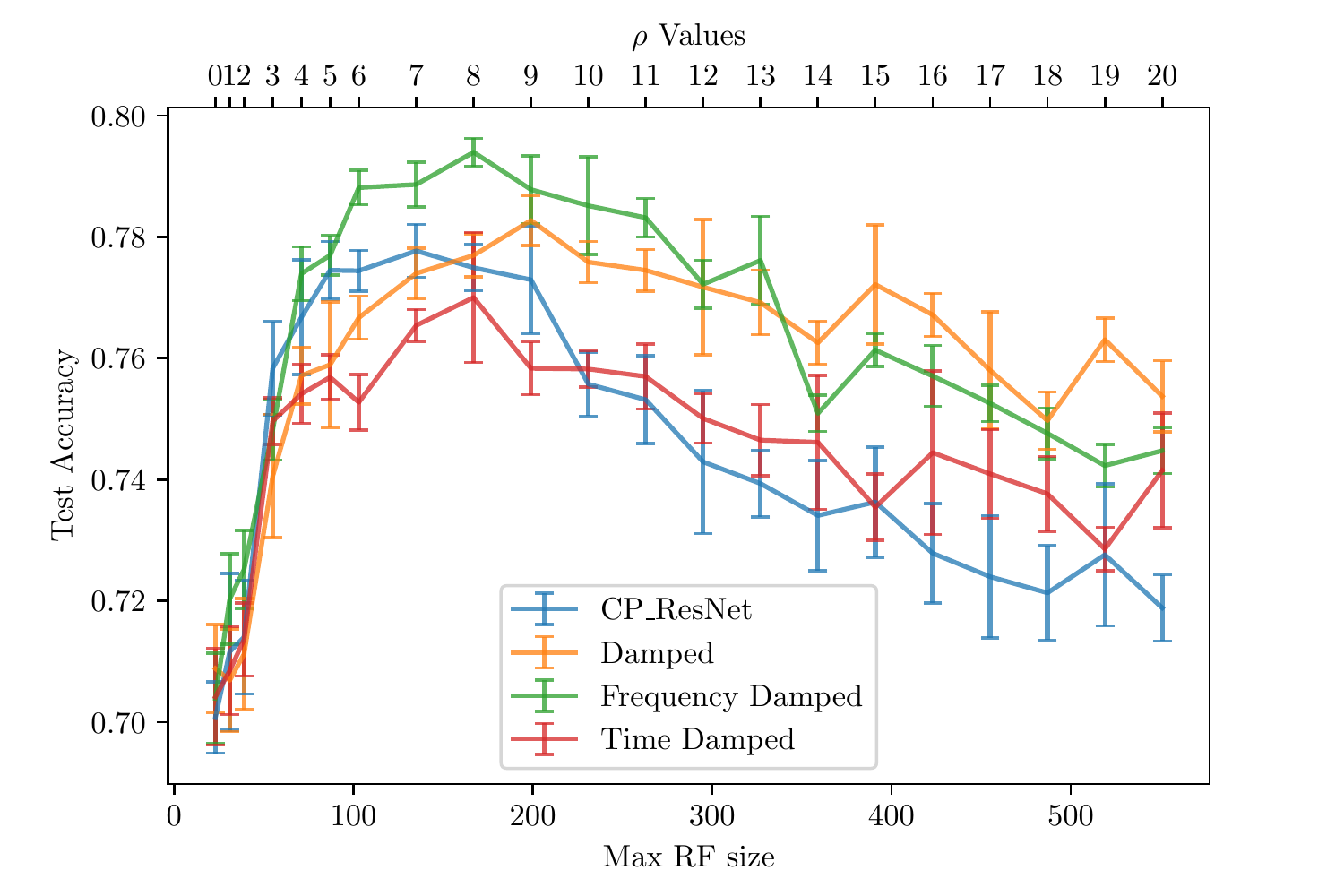}
\caption{Testing accuracy on DCASE`18 of CP\_ResNet 
and variants with damped filters over time dimension (Time~Damped), frequency dimension (Frequency~Damped) or both dimensions (Damped). The 4 networks have the same Max RF for a $\rho$ value but differ by the ERF size.}
\label{fig:acc_damped_vs_classic}
\end{figure}
\begin{figure}[t]
\centering
\includegraphics[width=3.7in]{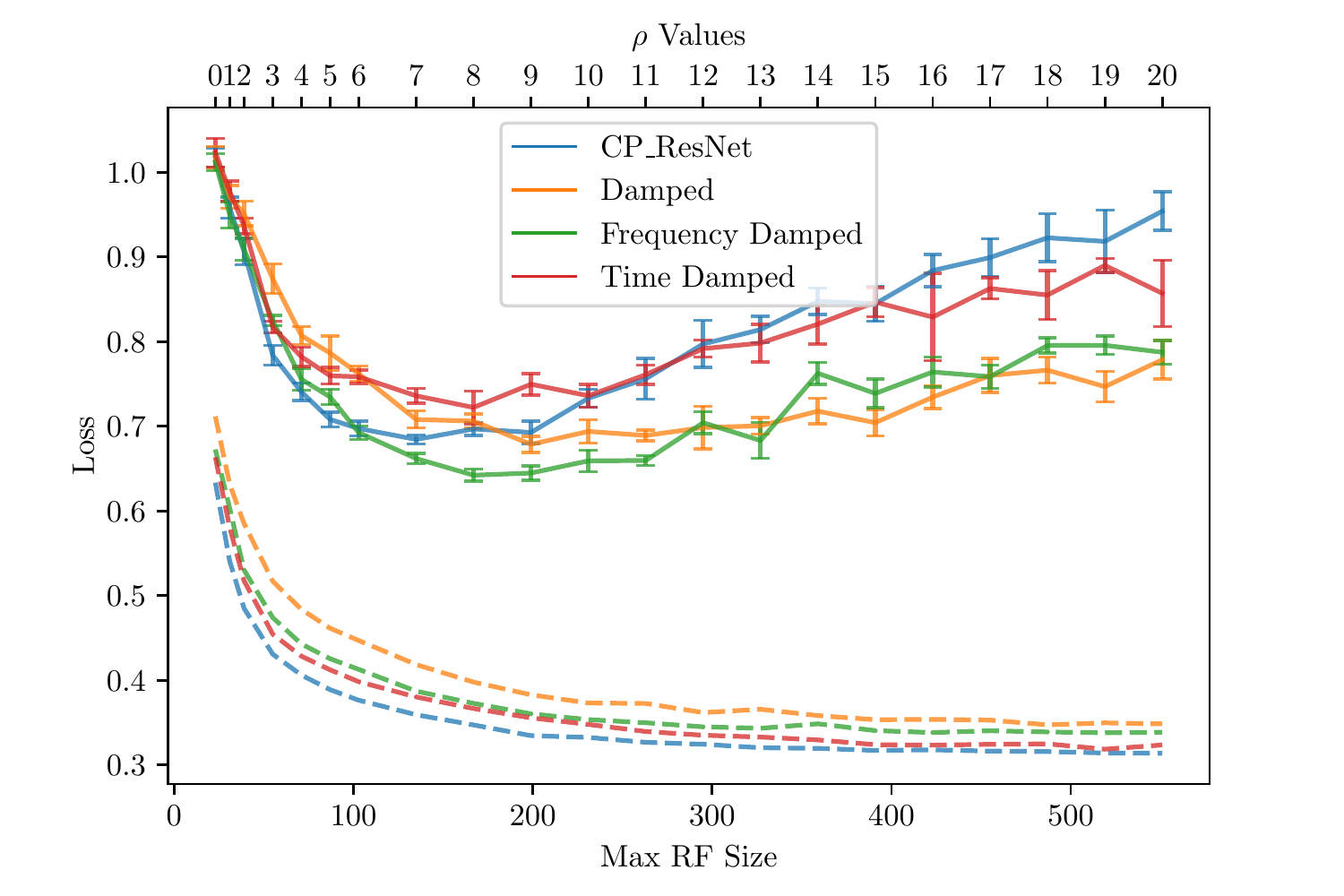}
\caption{Comparing Testing loss (solid lines) and training loss (dashed lines) on DCASE`18   of CP\_ResNet and variants with damped filters. The 4 networks have the same Max RF for a $\rho$ value but differ by the ERF size.}
\label{fig:loss_damped_vs_classic}

\end{figure}
\begin{figure}[t]
\centering
\includegraphics[width=3.5in]{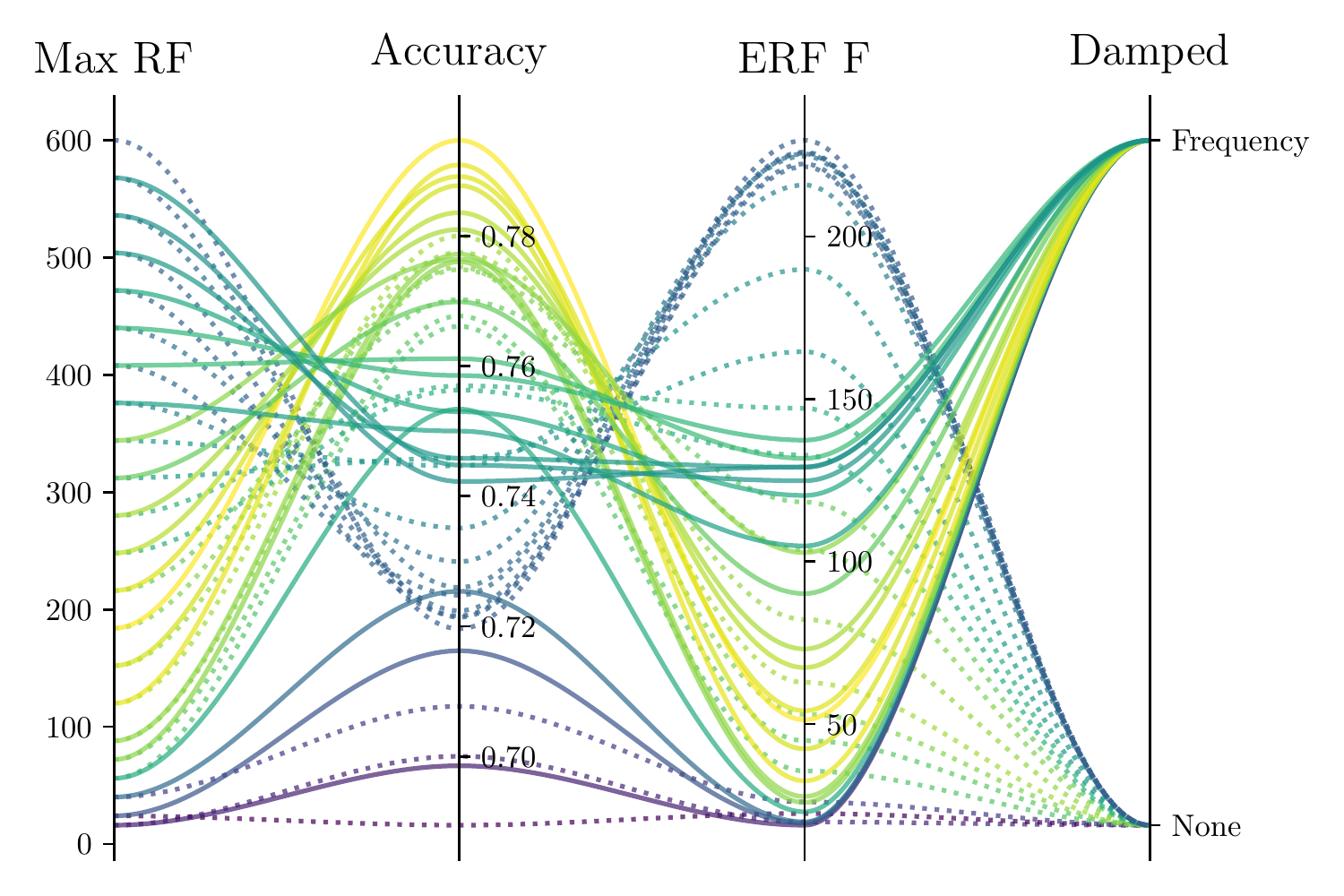}
\caption{Comparing the ERF over frequency (ERF F) and accuracy on DCASE`18 of a CP\_ResNet with damped filters over the frequency dimension (solid lines) vs with no damping (dashed lines). The color of the lines indicates the accuracy as can be read from the accuracy axis. 
}
\label{fig:par_damped_vs_classic}
\end{figure}

In this section, we present the results of our experiments with the damped CP\_ResNets. 
We calculate the Max RF size as explained in Section~\ref{sec:rf_cnns:mrf}, and additionally, calculate the ERF size as explained in Section~\ref{sec:rf_cnns:erf}.

 
Figures~\ref{fig:acc_damped_vs_classic} and~\ref{fig:loss_damped_vs_classic} show the influence of restricting the ERF by damping, on the performance of the CNN. 
As can be seen, the damped variants perform better than the non-damped ones as the Max RF grows larger. 
Additionally, the ``Frequency-Damped'' CNNs significantly outperform the original network on this task, indicating the importance of the inductive bias introduced by damping.  
This also indicates that restricting the RF (both ERF and Max RF) over the frequency dimension has a higher impact on generalization, for the task of ASC.
Figure~\ref{fig:loss_damped_vs_classic} shows that the damped networks have a larger loss on the training data but generalize better on the test data.

Figure~\ref{fig:par_damped_vs_classic} shows the correlation between the accuracy and the ERF size over frequency (ERF F) for both CP\_ResNet and  CP\_ResNet with damped filters over the frequency dimension.
While both networks have identical Max RF for a given configuration ($\rho$ value), 
damping helps in \emph{restricting} the ERF (over the frequency dimension) to the optimal range. In larger Max RF settings, the ERF size of the frequency damped version is significantly smaller, and therefore a larger range of Max RF settings (50, 450) have the optimal ERF achieving a minimum accuracy of $.76$, compared to the Max RF range of (50, 200) in the non-damped case.
Thus, filter damping can be considered as a simple but powerful tool for improving generalization on this task.


\begin{figure}[t]
\centering
\includegraphics[width=3.5in]{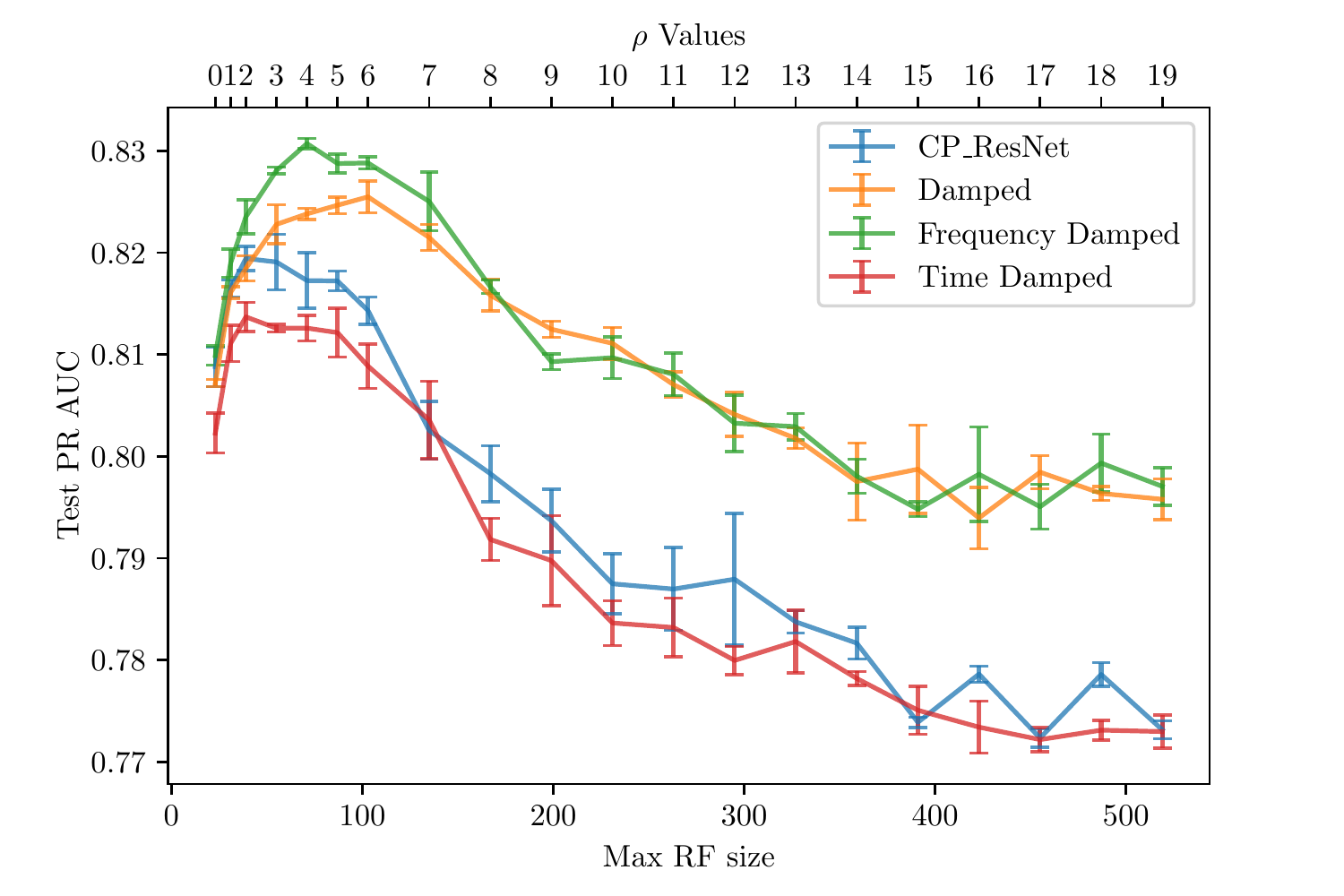}
\caption{PR-AUC  of CP\_ResNet and the damped variants  on  OpenMIC  dataset.  The Max  RF  size  of  the  network  over  both  time  and  frequency  dimensions. 
}
\label{fig:openmic_damped_vs_classic}
\end{figure}

\begin{figure}[t]
\centering
\includegraphics[width=3.5in]{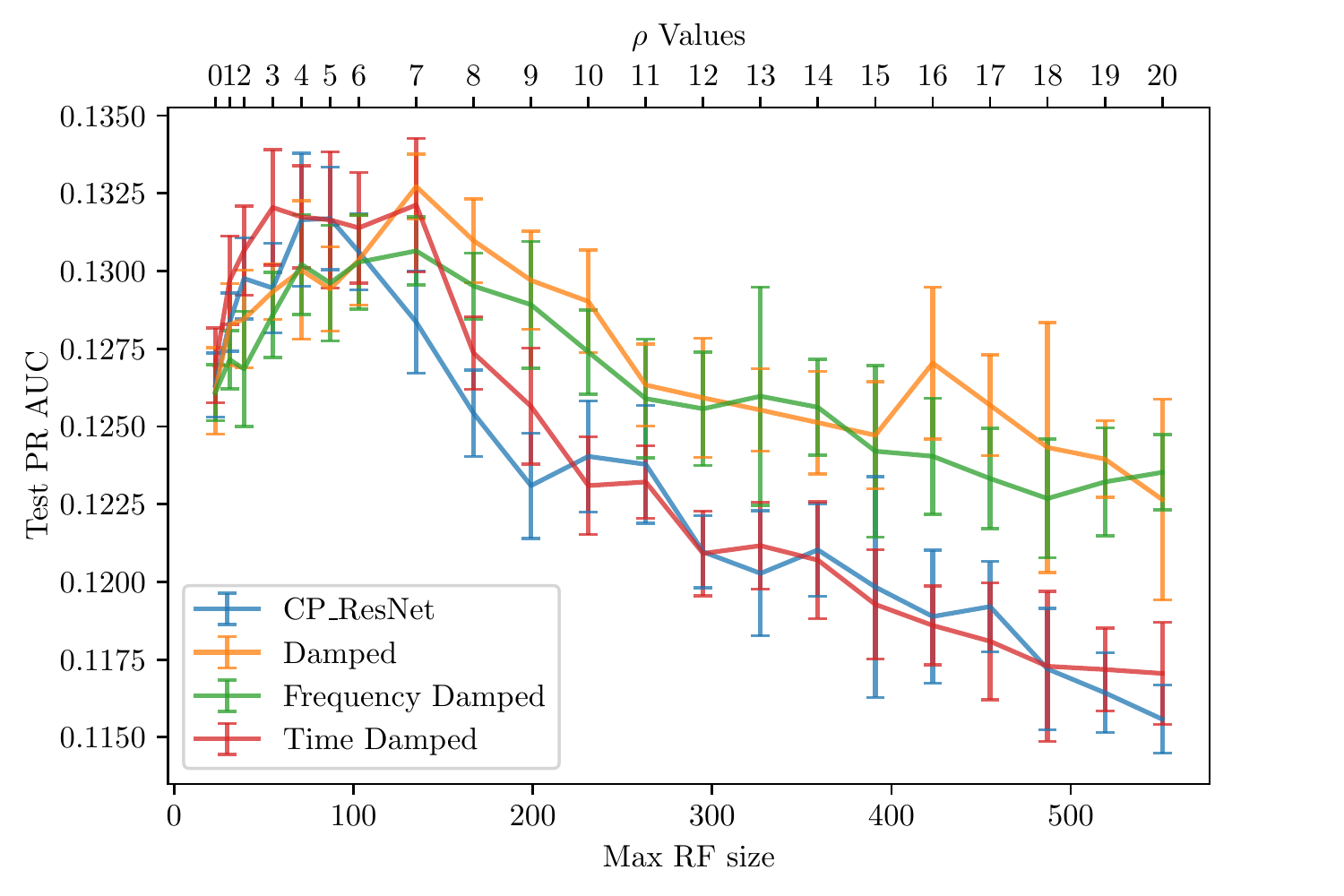}
\caption{PR-AUC  of CP\_ResNet and the damped variants  on  MediaEval  dataset.  The Max  RF  size  of  the  network  over  both  time  and  frequency  dimensions. 
} 
\label{fig:mediaeval_damped_vs_classic}
\end{figure}

Figure~\ref{fig:openmic_damped_vs_classic} shows a similar pattern on OpenMic dataset; damping over both dimensions and over the frequency dimension reduce overfitting as the Max RF increases. Furthermore, frequency damping boosts the performance of CP\_ResNet$_{\rho=4}$ to the state-of-the-art performance of  SS CP\_ResNet$_{\rho=7}$. 
The same trend can also be seen in  Figure~\ref{fig:mediaeval_damped_vs_classic} on the MediaEval-MTG-Jamendo Dataset.

\subsection{Relationship between RF-Regularization and the Number of Parameters}



Many factors influence the ability of CNNs to generalize, such as width and depth~\cite{TanL19effcientnet}, network architecture, and the number of parameters~\cite{NakkiranKBYBS20ICLR}.
Changing any of these will have an effect on the capacity of the network to fit the training data and to generalize to unseen data.

In Figures~\ref{fig:acc_damped_vs_classic},~\ref{fig:openmic_damped_vs_classic} and ~\ref{fig:mediaeval_damped_vs_classic}, for every $\rho$ value, CP\_ResNet and its three damped variants have the same number of parameters and the same topology.
The performance improvement can be explained by the difference in the size of the ERF between the four models. We influence the size of the ERF by introducing damping. 

 Both Figures~\ref{fig:train_vs_test_by_maxrf} and~\ref{fig:train_vs_test_by_maxrf_pooling} show the same pattern when comparing the training and testing loss of ResNets. However, in Figure~\ref{fig:train_vs_test_by_maxrf_pooling} all the networks have the same number of parameters; they only differ by the number of pooling layers.

\begin{figure}[t]
\centering
\includegraphics[width=3.5in]{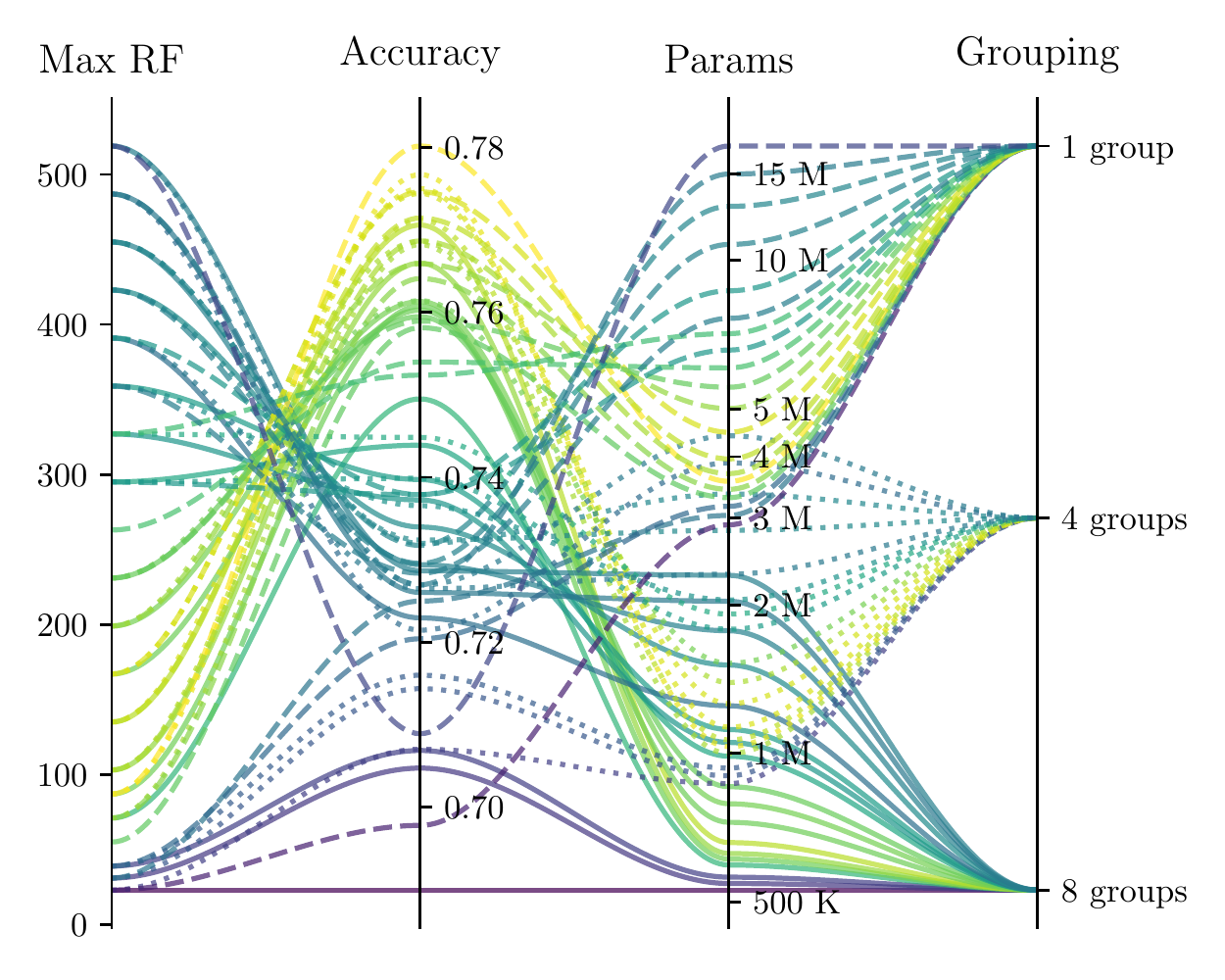}
\caption{Comparing the accuracy, number of parameters, and Max RF variants of CP\_ResNet with grouped convolutional layers on DCASE`18.
}
\label{fig:parallel_group_param}
\end{figure}

The max RF regularization method (as proposed in Section~\ref{sec:control_max_rf}) works  by changing the filter sizes to restrict the RF of the CNN. Changing the filter sizes of a \emph{convolutional layer} from $3 \times 3$ to $1 \times 1$ does change the number of parameters of the \emph{layer} from $3 \times 3 \times C_{in} \times C_{out}$ to $1 \times 1 \times C_{in} \times C_{out}$, where $C_{in}$, $C_{out}$ are the number of input and output channels respectively.
In order to disentangle the number of parameters from the Max RF, we use grouped convolutional layers~\cite{xie2017aggregated}. Using grouped convolution does not change the max RF of the network, but rather changes the RF of a layer over its input channels. In other words, grouped convolutional layers can be seen as: first, slicing the input over the channels dimension to $g$ groups, each with $C_{in}/g$ channels;
second, applying the convolution operator on each group resulting in $C_{out}/g$ each; and finally, concatenating the output of the $g$ groups over the channels dimension, resulting in $C_{out}$ channels. As a result, it changes the number of parameter of \emph{every layer} from  $ X_t \times X_f \times C_{in} \times C_{out}$ to  $ g \times X_t \times X_f \times (C_{in}/g) \times C_{out}/g = X_t \times X_f \times C_{in} \times C_{out}/g$, where $X_t$,$X_f$ are the filter size over time and frequency respectively, and $g$ is the number of groups. In short, grouping is a simple method to reduce the number of parameters without affecting the spatial max RF of the CNN.
Figure~\ref{fig:parallel_group_param} compares the accuracy on DCASE`18 and number of parameters of CP\_ResNet with variants with grouped convolutional layers using 4 groups and 8 groups. The figure shows that the accuracy of the networks is correlated with the RF and not with the number of parameters.




%% file: s9_conclusion.tex
\label{sec:conclusion}

In this paper, we have analyzed the impact of the maximum and the effective receptive field on the generalization capability of deep convolutional neural networks for various audio classification and tagging tasks. 
We investigated the influence of the receptive field on both the time dimension and the frequency dimension. 
We showed that the size of the receptive field (especially over the frequency dimension) is crucial for the generalization of CNNs on audio tagging and classification tasks, in contrast to computer vision where the common belief is that deeper networks (and thus larger RF) generalize better.
We demonstrated this on different tasks, and for different CNN architectures.

Based on these findings, we introduced the receptive field size as a new and important hyper-parameter that researchers should take into account when designing and tuning CNN architectures for various audio classification and tagging tasks. 
We proposed a general method for controlling the Max RF and applied it to widely-used CNNs, resulting in CNN families whose Max RF can be controlled with the introduced hyper-parameter. 
We also proposed filter damping as a method to control the Effective RF of a CNN, without the need for altering the architecture. We empirically showed that RF-regularized CNNs achieve state-of-the-art performance on three diverse audio classification and tagging tasks, outperforming much more complex approaches.

We conclude that RF-regularization is a simple but general technique that can be adopted to improve the generalization capabilities in various CNN architectures, and across many tasks, without the use of any additional training data, pre-training, or complex computational modules.

%% file: a0_exp_setup.tex
\label{app:exp_setup}

In this section, we explain the technical details and setup we use to perform the experiments reported throughout the paper. %

\subsubsection{Spectrograms}
We use Short Time Fourier
Transform (STFT) with a window size of 2048 to extract the CNNs input features from the audio clips. We use 25\% overlap between the STFT windows in the Emotion and theme tagging task because the input songs are longer than the audio clips in the other tasks, where we 75\% overlap.  We apply perceptual weighting on the resulting spectrograms and apply a mel-scaled filter bank in a similar fashion to Dorfer et al.~\cite{DorferDCASE2018task1}. This preprocessing results in 256 mel frequency bins. The RF sizes reported in this paper are dependent on the resolution of the spectrograms, especially--as we have shown--over the frequency dimension.  
We use the training set mean and standard deviation to normalize the input spectrograms.

\subsubsection{Augmentation}
We use  \textit{Mix-up}~\cite{zhangMixupEmpiricalRisk2017}, which works by linearly combining two input samples and their targets. It was shown to be an effective augmentation method that is simple but can have a great impact on the performance and the generalization of the models in tasks~\cite{DorferDCASE2018task1,Koutini2019Receptive}.
In our experiments, we set $\lambda=0.3$ for ASC and emotion and theme recognition tasks. We do not use Mix-up on the OpenMIC dataset because our preliminary experiments showed that Mix-up does not help in this task.

We also roll the spectrograms randomly over the time dimension. More precisely, we roll the input spectrograms of each clip by a number of pixels sampled uniformly from the range $(-50,50)$.

\subsubsection{Optimization}

In a setup similar to ~\cite{Koutini2019Receptive,Koutinitrrfcnns2019,koutinifaresnet2019}, we use Adam~\cite{kingmaAdamMethodStochastic2014} to train our models. 
We use a linear learning rate scheduler for 100 epochs, dropping the learning rate from $1 \times 10^{-4}$ to $1 \times 10^{-6}$. 
In the end, we train for 40 additional epochs using the minimum learning rate of $1 \times 10^{-6}$.
We report the mean and the standard deviation of 3 runs.

%% file: a1_arch.tex
\label{app:arch:design:components:maxrf}

Equation~\ref{eq:calc_maxrf} in Section~\ref{sec:rf_cnns} provides a method to calculate the Max RF of a layer based on the Max RF of the previous layer, the accumulative stride, the filter size, and the stride of the respected layer. 
Araujo et al.~\cite{araujo2019computingreceptivefield} show a detailed method to compute the Max RF of a CNN. 
As they show, many factors play a role in the Max RF of CNN. 
For instance, two very similar architectures with few differences -- such as introducing a single sub-sampling layer -- can have a very different max RF. 
In this section, we highlight the common building components of CNNs and how they change the RF of the network.

\subsubsection{Filter Sizes}
Filter sizes affect directly the Max RF of a layer as they determine the number of spatial pixels of the preceding layer (or input) that can affect the activation of this layer. For instance, in Section~\ref{sec:control_maxrf:resnet}, we use the filter sizes to have granular control of the Max RF of CNNs.

\subsubsection{Strides}
Increasing the stride has a high impact on the Max RF of a CNN as shown in Equation~\ref{eq:calc_maxrf}, since the stride of a layer is multiplied by the accumulated stride with respect to the input. 
A stride of 2 means that two spatially adjacent pixels of a layer output cover double the pixels of the preceding layer compared to the case with a stride of 1.

\subsubsection{Pooling layers}
\begin{figure}[h]
\centering
\includegraphics[width=3.5in]{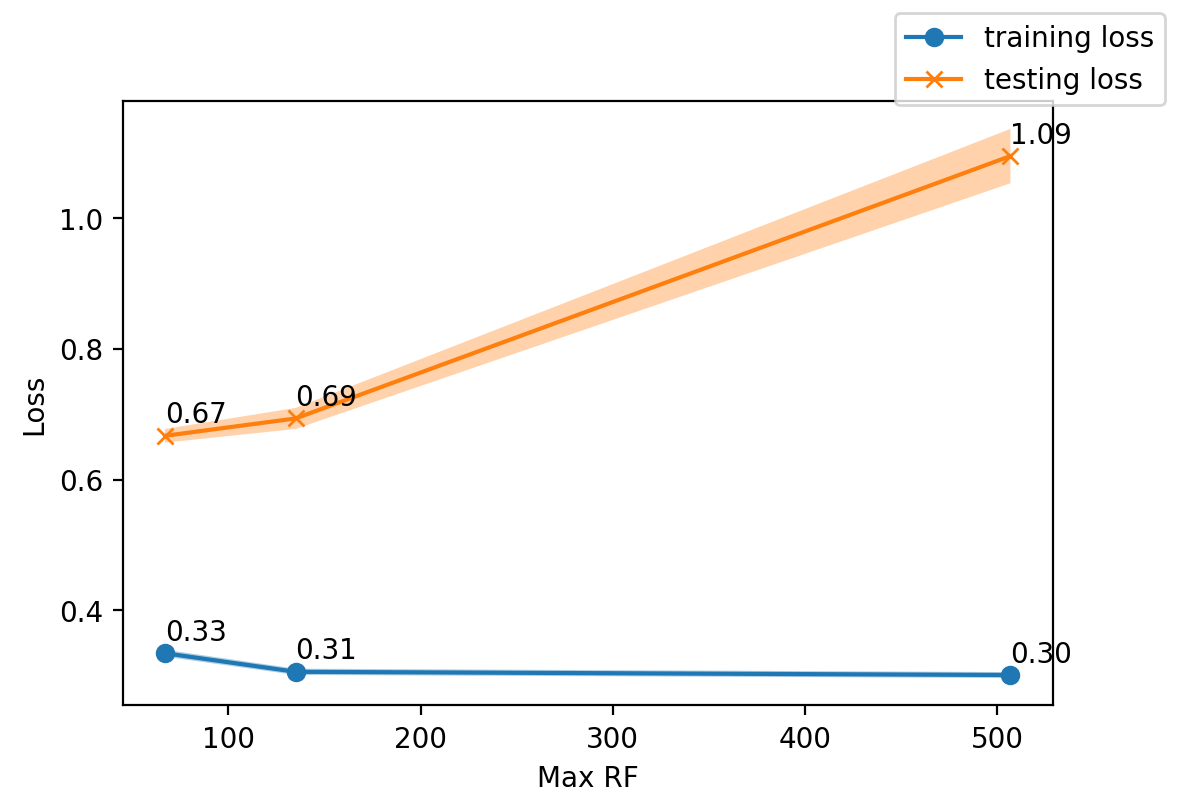}
\caption{Comparing the training loss vs the testing loss of ResNets with the same number of parameters but with different Max RF. The difference in RF size is due to adding pooling layers. The increase of the testing loss with larger Max RF indicates overfitting. The CNNs are trained on DCASE`18.}
\label{fig:train_vs_test_by_maxrf_pooling}
\end{figure}

Pooling layers can be seen as standard convolutional layers, but with no learnable parameters. 
We can use pooling layers to increase the Max RF without changing the number of parameters.
Figure~\ref{fig:train_vs_test_by_maxrf_pooling} compares the training and testing loss of ResNets with the same number of parameters, but with different Max RF. We change the RF of these networks by introducing $2 \times 2$ pooling  layers. We then calculate the Max RF, train the network, and report the training and testing loss. Figure~\ref{fig:train_vs_test_by_maxrf_pooling} shows a similar pattern to Figure~\ref{fig:train_vs_test_by_maxrf}: the training loss drops as the RF grows, but the testing loss increases.

\subsubsection{Skip and Residual Connections}
Skip and residual connections are very common in modern architectures such as ResNet~\cite{heDeepResidualLearning2016} and DenseNet~\cite{huangDenselyConnectedConvolutional2017}. Since we are studying the Max RF, if two branches are added or concatenated as an input for a layer, we consider the branch with the bigger Max RF when calculating the Max RF of the layer.
Figure~\ref{fig:dcase18:dilated:grouped:growth:acc} shows the testing accuracy of variants of DenseNets with different  \emph{growth rate}. The growth rate determines the number of channels in the output of each Dense layer. This output is then presented as an input to all the following Dense layers, which is known as a \emph{Skip Connection}. We show that all the variants degrade in performance in larger RF settings.

\subsubsection{Dilated Convolution}
\label{sec:sub:dilated:cnn}
A dilated convolutional layer~\cite{YuK15DilatedConvolutions} can be seen as a convolutional layer with a larger sparse filters which has the same number of learnable parameters. 
The effective filter size of dilated convolutional layer with $d_n$ dilation is $ k_n^* = d_n(k_n-1)+1$. When $d=1$, $k_n^*=k_n$. However, when $d_n>1$, for receptive field calculation purposes, the dilated convolutional layer (with dilation $d_n$ and filter size $k_n$) can be seen as a regular convolutional with a larger filter $ k_n^*$. Therefore, we can replace $k_n$ with $ k_n^*$ in Equation~\ref{eq:calc_maxrf} to calculate its receptive field. 

In the case of $d_n>1$, although the activation of a spatial pixel of the output of a layer is affected by the same number of spatial pixels of the preceding layer as with $d=1$, two adjacent pixels cover a bigger number of pixels of the preceding layer. 

\begin{figure}[h]
\centering
\includegraphics[width=3.5in]{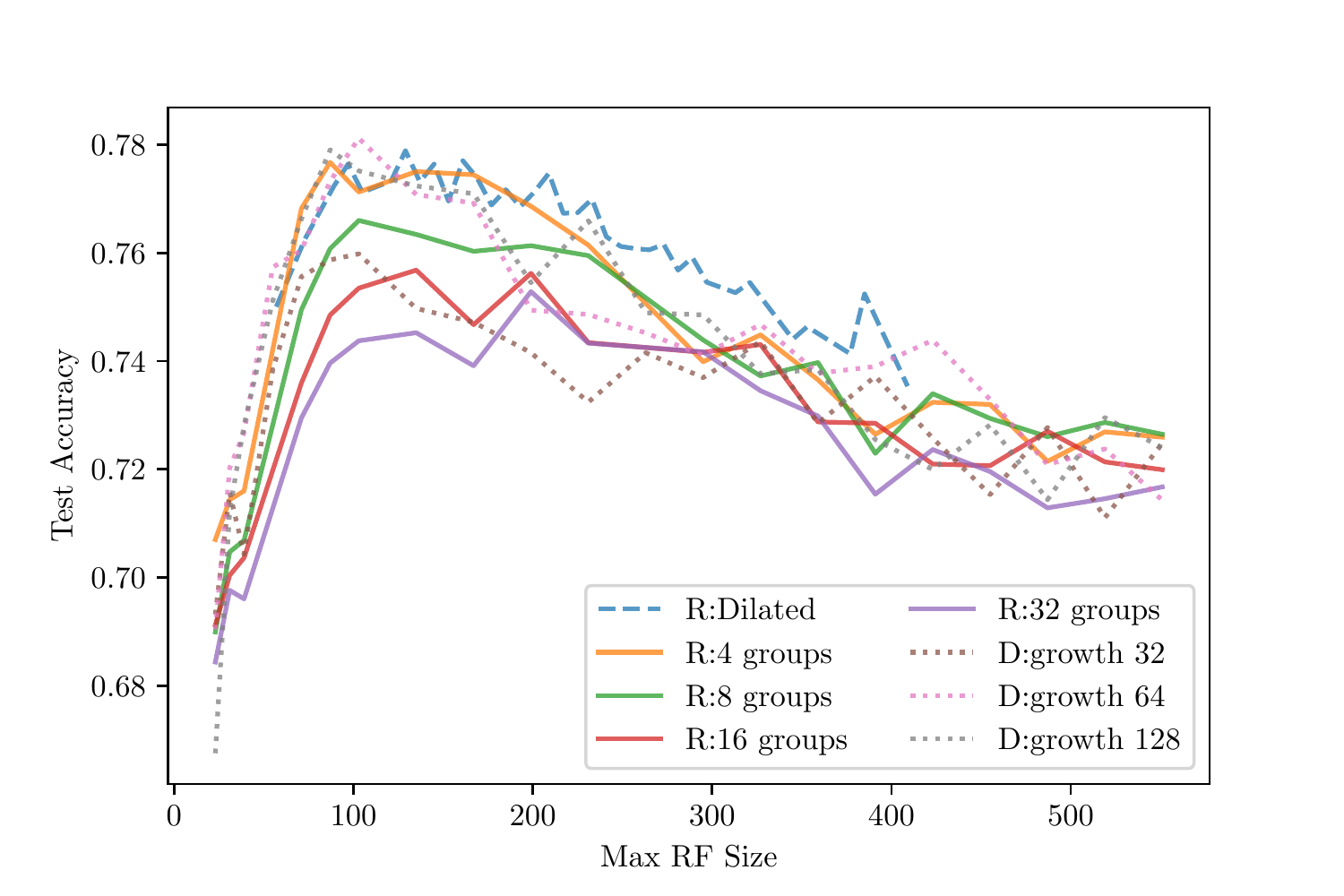}
\caption{  DCASE`18  testing accuracy of different designs of CP\_ResNet (R) and  CP\_DenseNet (D) convolutional layers. }
\label{fig:dcase18:dilated:grouped:growth:acc}
\end{figure}

Figure~\ref{fig:dcase18:dilated:grouped:growth:acc} shows the testing accuracy of a variants of CP\_ResNet with dilated convolutional layers.
We start the experiment with variants of CP\_ResNet with a Max RF ranging from $55 \time 55$ pixels to $199 \time 199$ (corresponding $\rho$ from $3$ to $9$). 
We then change the convolution of different layers (in different experiments) to a dilated convolutional operator with $d=2$. 
We then calculate the Max RF and train the network and report the testing accuracy. 

\subsubsection{Grouped Convolution}
Grouped convolution restricts the receptive field of a layer over the channels dimension, and therefore does not impact the spatial RF of a CNN. 
This indicates that the same approach to control the Max RF can be applied to architectures that utilize grouped convolutional layers such as ResNext~\cite{xie2017aggregated}.
Figure~\ref{fig:dcase18:dilated:grouped:growth:acc} shows that the performance of variants of CP\_ResNet with a different number of groups follows the same pattern.

\subsubsection{Batch Normalization}

Batch Normalization~\cite{ioffe2015batch} in CNNs is calculated per channel. 
However, the mean and the variance are calculated across all the spatial pixels during training. During inference, the layer uses the training statistics and thus does not affect the Max RF.

\subsubsection{Shake-Shake}

Shake-Shake~\cite{gastaldi2017shake} regularization works by replacing the summation operation of residual branches by a stochastic affine combination. 
Gastaldi~\cite{gastaldi2017shake} shows empirically that using this regularization technique improves the generalization of a ResNet with two residual branches on various computer vision datasets. 
Gastaldi also shows that using different scaling coefficient for the residual branches in the backward pass improves the generalization. 
Previous work~\cite{Koutinitrrfcnns2019,koutini2019emotion} confirms that Shake-Shake regularization can  boost performance on unseen data in some tasks as well, especially in the presence of noisy data or labels. 

In Shake-Shake regularized architectures, both of the residual branches will have the same Max RF. 
Therefore, combining both with different scaling factors does not change the Max RF of the network. 
We apply Shake-Shake regularization to \emph{CP\_ResNet} in our experiments. The resulting network (SS CP\_ResNet) will have consequently the same Max RF as explained in Section~\ref{sec:control_maxrf:resnet}.

\subsubsection{Frequency-Aware CNNs}

Frequency-Aware CNNs~\cite{koutinifaresnet2019} (FACNNs) showed empirically improvements in the performance of CNNs in acoustic scene classification~\mbox{\cite{koutinifaresnet2019}}, and emotion and theme detection~\mbox{\cite{koutini2019emotion}} tasks. 
Frequency-aware convolutional layers work by concatenating a new channel to the input,
that provides a positional encoding for the frequency.
Unlike CNNs with restricted Max RF, CNNs with a large RF can infer the frequency positional information of their input.
Hence, FACNNs were proposed to compensate for the lack of frequency-range information in CNNs with a restricted RF over the frequency dimension.  
Since adding the frequency information as a new channel does not affect the Max RF of the CNN, \emph{CP\_ResNet} with frequency-aware layers (\emph{CP\_FAResNet}) will have the same Max RF. 






\begin{table}[]
\caption{Per-class accuracy Comparison of CP\_ResNet with the state of the art on the DCASE`19 Dataset}
\begin{center}

\begin{tabular}{l|l|l|l}
                  & Chen et al.\cite{chen2019dcaseasc} & CP\_ResNet CV & Seo et al.~\cite{Hyeji2019dcaseasc} \\ \hline
Overall           & 85.2 \%                           & 83.7 \%       & 82.5 \%      \\
\hline
Airport           & 77.5 \%                           & 79.9 \%       & 74.9 \%      \\
Bus               & 96.1 \%                           & 96.2 \%       & 95   \%      \\
Metro             & 89.9 \%                           & 84.7 \%       & 82.9 \%      \\
Metro station     & 82.5 \%                           & 81.5 \%       & 80   \%      \\
Park              & 96.7 \%                           & 96.1 \%       & 92.4 \%      \\
Public square     & 67.5 \%                           & 64   \%       & 60.7 \%      \\
Shopping mall     & 80   \%                           & 80.1 \%       & 85.4 \%      \\
Street pedestrian & 78.6 \%                           & 72.9 \%       & 68.6 \%      \\
Street traffic    & 92.8 \%                           & 92.6 \%       & 93.2 \%      \\
Tram              & 90.6 \%                           & 89   \%       & 91.9 \%     
\end{tabular}
\label{tab:res:asc:sota:perclass}
\end{center}
\end{table}